\documentclass{article}

\usepackage{arxiv}

\usepackage[utf8]{inputenc} 
\usepackage[T1]{fontenc}    
\usepackage{hyperref,url,booktabs,amsfonts,nicefrac,microtype,graphicx,doi,lineno,microtype,subcaption,amssymb,framed,tabularx,cleveref,xcolor,soul,multicol,array,tipa,xr,threeparttable,chngcntr,enumitem,etaremune}

\title{Artificial Intelligence Can Match Domain Experts in Evidence Extraction and Critical Appraisal of Microbial Oncogenesis Research Publications}

\author{ \href{https://orcid.org/0009-0002-9916-0049}{\includegraphics[scale=0.06]{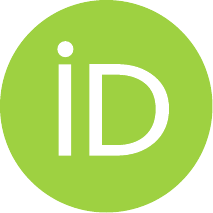}\hspace{1mm}Kaela Kokkas}$^{1}$\\   \texttt{2423686@students.wits.ac.za} \\
\And
\href{https://orcid.org/0000-0001-8770-5916}{\includegraphics[scale=0.06]{orcid.pdf}\hspace{1mm}Hairong Wang}$^{4,15}$ \\
\And
\href{https://orcid.org/0000-0003-0783-2072}{\includegraphics[scale=0.06]{orcid.pdf}\hspace{1mm}Richard Klein}$^{4,15}$ \\
\And
\href{https://orcid.org/0000-0001-7696-3776}{\includegraphics[scale=0.06]{orcid.pdf}\hspace{1mm}Nazir A. Ismail}$^{2}$ \\
\And
\href{https://orcid.org/0000-0002-6911-3615}{\includegraphics[scale=0.06]{orcid.pdf}\hspace{1mm}Natalie Irwin}$^{3,5}$ \\
\And
\href{https://orcid.org/0009-0006-5321-8179}{\includegraphics[scale=0.06]{orcid.pdf}\hspace{1mm}Mohammad Z. Moonsamy}$^{4,5}$ \\
\And
\href{https://orcid.org/0000-0003-2874-2703}{\includegraphics[scale=0.06]{orcid.pdf}\hspace{1mm}Kubendran Naidoo}$^{5,6,7,8,9,10}$ \\
\And
\href{https://orcid.org/0000-0003-4115-894X}{\includegraphics[scale=0.06]{orcid.pdf}\hspace{1mm}Jeremy Nel}$^{5,11}$ \\
\And
\href{https://orcid.org/0000-0003-4026-1697}{\includegraphics[scale=0.06]{orcid.pdf}\hspace{1mm}Ekene E. Nweke}$^{5,12}$ \\
\And
\href{https://orcid.org/0000-0002-5476-1011}{\includegraphics[scale=0.06]{orcid.pdf}\hspace{1mm}Raveen Parboosing}$^{13}$ \\
\And
\href{https://orcid.org/0000-0002-8231-6935}{\includegraphics[scale=0.06]{orcid.pdf}\hspace{1mm}Emmanuel K. Sekyi}$^{14}$ \\
\And
\href{https://orcid.org/0000-0002-1808-1588}{\includegraphics[scale=0.06]{orcid.pdf}\hspace{1mm}Rebecca T. van Dorsten}$^{5,6,7}$ \\
\And
\href{https://orcid.org/0000-0001-7700-1069}{\includegraphics[scale=0.06]{orcid.pdf}\hspace{1mm}Bruce A. Bassett}$^{4,5,15}$ \\
\And
\href{https://orcid.org/0000-0002-7099-2936}{\includegraphics[scale=0.06]{orcid.pdf}\hspace{1mm}Robert F. Breiman}$^{5,15,16}$ \\ }

\date{}
\renewcommand{\headeright}{}
\renewcommand{\undertitle}{}
\renewcommand{\shorttitle}{AI Can Match Domain Experts in Evidence Extraction and Appraisal}

\hypersetup{
pdftitle={Artificial Intelligence Can Match Domain Experts in Evidence Extraction and Critical Appraisal of Microbial Oncogenesis Research Publications},
pdfsubject={q-bio.QM, cs.CL, cs.AI},
pdfauthor={Kaela~Kokkas, Hairong~Wang, Richard~Klein, Nazir A.~Ismail, Natalie~Irwin, Mohammad Z.~Moonsamy, Kubendran~Naidoo, Jeremy~Nel, Ekene E.~Nweke, Raveen~Parboosing, Emmanuel K.~Sekyi, Rebecca T.~van~Dorsten, Bruce A.~Bassett, Robert F.~Breiman},
pdfkeywords={large language models, artificial intelligence, critical appraisal, evidence extraction, evidence synthesis, evidence evaluation, biomedical literature, microbial oncogenesis}
}

\begin{document}
\maketitle

    $^{1}$Department of Clinical Microbiology and Infectious Diseases, Faculty of Health Sciences, University of the Witwatersrand, Johannesburg, South Africa \\
    $^{2}$ Department of Clinical Microbiology and Infectious Diseases, National Health Laboratory Service and Faculty of Health Sciences, University of the Witwatersrand, Johannesburg, South Africa \\
    $^{3}$ Division of Medical Oncology, Department of Internal Medicine, University of the Witwatersrand, Johannesburg, South Africa \\
    $^{4}$ School of Computer Science and Applied Mathematics, University of the Witwatersrand, Johannesburg, South Africa \\
    $^{5}$ Infectious Diseases and Oncology Research Institute (IDORI), Faculty of Health Sciences, University of the Witwatersrand, Johannesburg, South Africa \\
    $^{6}$ South African Medical Research Council Vaccines and Infectious Diseases Analytics Research Unit, Faculty of Health Sciences, University of the Witwatersrand, Johannesburg, South Africa \\
    $^{7}$ South African Medical Research Council Wits Antiviral Gene Therapy Research Unit, Faculty of Health Sciences, University of the Witwatersrand, Johannesburg, South Africa\\
    $^{8}$ National Health Laboratory Service, Johannesburg, South Africa\\
    $^{9}$ Department of Molecular Medicine and Haematology, School of Pathology, University of the Witwatersrand, Johannesburg, South Africa\\
    $^{10}$ Wits Research Institute for Malaria, Faculty of Health Sciences, National Health Laboratory Service, University of the Witwatersrand, Johannesburg, South Africa\\
    $^{11}$ Division of Infectious Diseases, School of Clinical Medicine, Faculty of Health Sciences, University of the Witwatersrand, Johannesburg, South Africa \\
    $^{12}$ Department of Surgery, Faculty of Health Sciences, University of the Witwatersrand, Johannesburg, South Africa\\
    $^{13}$ Division of Virology, University of the Witwatersrand \& National Health Laboratory Service, Johannesburg, South Africa \\
    $^{14}$ OncoVectra, London, United Kingdom\\
    $^{15}$ Wits Machine Intelligence and Neural Discovery (MIND) Institute, University of the Witwatersrand, Johannesburg, South Africa \\
    $^{16}$ Department of Global Health, Rollins School of Public Health, Emory University, Atlanta, GA United States

\vspace{15em}

\begin{abstract}
\textbf{Background}: Confirmed oncogenic microbes contribute significantly to cancer burden. Identifying and confirming novel microbial oncogenicity could yield strategies and tools that will reduce disease burdens. However, relevant evidence may be dispersed across a vast biomedical literature that is infeasible for humans to comprehensively synthesize. Large Language Models (LLMs) may enable scalable, expert-level systematic evidence synthesis to identify high priority microbe-cancer pairs; however, such capabilities have not yet been demonstrated.\newline
\textbf{Methods}: Domain experts were recruited to create a human-validated test dataset to benchmark the performance of LLMs (Gemini 2.5 Pro, Gemini 2.5 Flash, GPT-5, and GPT-5 Nano) on 24 original research papers using Mouse Mammary Tumor Virus-Like Virus and breast cancer as a case study. We devised a structured template for evidence extraction and appraisal of papers, consisting of multiple choice, Likert-scale, multi-select, and free-text question types (77 question items across 24 papers). Agreement between (1) experts, and (2) experts and each LLM, was determined per question instance using novel scoring metrics. LLMs were assessed by comparing inter-expert and expert-LLM agreement score distributions to determine whether LLMs behaved as additional experts by either increasing or maintaining inter-expert agreement. Free-text responses were further evaluated qualitatively. \newline
\textbf{Results}: Across all question types, LLM responses aligned closely with expert assessments, with two models (GPT-5, GPT-5 Nano) achieving score distributions statistically indistinguishable from those of experts. Gemini models behaved similarly for most tasks but were significantly more lenient in applying microbial oncogenesis criteria, often over-attributing criteria fulfillment. Hallucinations were rare, although more frequent in smaller models (Gemini 2.5 Flash, GPT-5 Nano). Methodological appraisal and identification of contradictions within full-text papers were the most persistent areas of LLM vulnerability, however, the error rate could not be directly compared with experts. \newline
\textbf{Conclusions}: Two LLMs (GPT-5, GPT-5 Nano) were indistinguishable from domain experts on structured domain research paper evaluation tasks. This evidence supports use of LLMs for automated systematic evidence synthesis. However, methodological appraisal tasks and contradiction identification in full-text papers remain weaknesses requiring further investigation, strengthening, and possibly multi-model strategies.
\end{abstract}


\section{Introduction}
The identification of human papillomavirus (HPV) as the cause of cervical cancer \cite{Watts2023} established cancer as a preventable disease through vaccination, screening, and treatment of infection \cite{WorldHealthOrganisationWHO2022}. Since then, 11 microbial Group 1 carcinogens (refers to those agents with sufficient evidence of carcinogenicity in humans) have been identified \cite{InternationalAgencyforResearchonCancerIARC2025, IARC2012}, with infectious agents linked to at least 20\% of cancer cases worldwide and 30\% of cases in sub-Saharan Africa \cite{deMartel2020, Parkin2020}. The identification of additional oncogenic microbes could provide opportunities to significantly reduce cancer incidence, morbidity, and mortality via development and integration of targeted interventions into public health policies.

However, identifying the most probable and impactful microbe-cancer pairs (MCPs) for further research is a significant challenge, with over 1400 possible microbial species \cite{NatureReviewsMicrobiology2011}, more than 100 cancer types \cite{NationalCancerInstitute2021, NationalCancerInstituteAZ}, and variability in oncogenic mechanisms, the microbial types, and their molecular structures which could impact oncogenicity \cite{Moore2010}. This has resulted in a division of efforts with multiple MCPs being investigated simultaneously \cite{Luczynski2022, Miyabayashi2022, Bernardo2023, BouZerdan2022, Dougherty2023}. While the existing literature exploring MCPs is not exhaustive, there are many proposed pairs with compelling evidence that warrant further investigation. Identifying a starting point requires a comprehensive, systematic evidence synthesis process to determine which MCPs offer both the greatest plausibility and public health benefit. However, the existing literature and one million articles published yearly in more than 13 000 active indexed biomedical science journals \cite{Ghasemi2022}, renders a traditional ``human-driven'' literature search and evaluation infeasible, even after screening and filtering for relevance. Artificial Intelligence (AI) provides a potential for augmenting traditional literature search strategies with tools capable of rapidly evaluating and synthesizing available evidence.

AI, especially in the form of Large Language Models (LLMs), is increasingly demonstrating value for strengthening and accelerating a variety of facets of medical research, from reviewing and evaluating medical data \cite{Koh2025, Blbl2025, Brigo2025, Gottweis2026}, drug discovery \cite{Zhang2025}, predicting and designing protein folding structures \cite{Jumper2021, Zambaldi2024}, to multi-use research assistants \cite{Chappell2023, Nordmann2025, vandeSchoot2021, Li2025, Tsafnat2014, Huang2025}, and specialized systemic review tools. Recent advancements in accessibility, affordability, and the capability of pre-trained LLMs, including the advent of reasoning LLMs capable of solving complex reasoning tasks, make analyzing the vast scientific literature potentially feasible. However, concerns regarding the trustworthiness of these tools necessitate validation \cite{McFadden2023}. These concerns include: (1) LLMs’ tendency to hallucinate and produce plausibly sounding yet factually incorrect outputs \cite{Kim2025}, and (2) limited assessments of the validity of outputs complicated by the ``black-box problem'' where neural networks’ rationale and critical thinking processes are indiscernible \cite{Castelvecchi2016}. Thus, to allow for further integration of these systems into current research workflows, we need to ensure that models’ outputs are consistently trustworthy, accurate, and controllable when performing these tasks. 

\begin{figure}[!htb]
    \centering
        \caption{Overview of proposed AI system. Initial keyword search is performed for microbe-cancer pairs (MCPs) on research databases, with web scraping and classification of titles and abstracts by small LLMs using pre-defined relevancy criteria. Papers classified as \textit{Somewhat relevant} may be manually reviewed or included in next steps if limited papers retrieved. Full-text versions of \textit{Relevant} papers are downloaded and analyzed by LLMs using extraction template. Potentially \textit{Relevant} papers in references are cross-checked with paper storage database and searched for if absent. LLMs produce final MCP evaluation using individual paper summaries and data from infection and cancer registries to determine geographical impact and prevalence of both the cancer and the microbe (when results are available and accessible). This final MCP evaluation is then added to a dynamic list of MCPs, ranked based on a standardized scoring matrix including factors such as potential impact if targeted, confidence in causal relationship, and novelty, and updated with each new MCP evaluation.}
    \label{fig:proposed_framework}
    \includegraphics[width=0.9\linewidth]{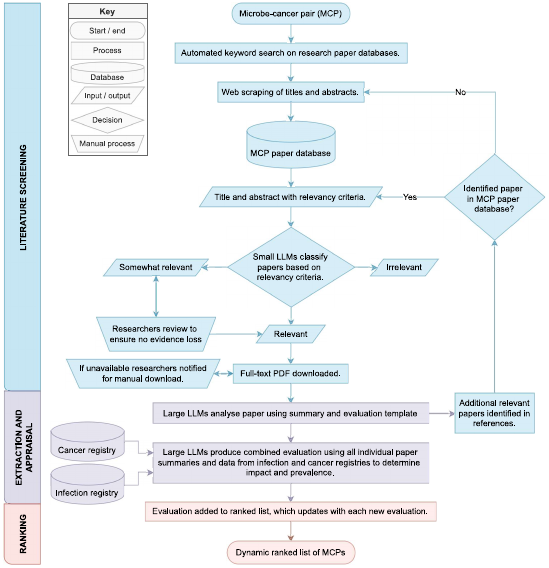}
\end{figure}

We envision using AI to filter and analyze literature on MCPs, ultimately producing a ranked list based on a standardized scoring matrix for prioritizing investigations (Fig.~\ref{fig:proposed_framework}). To begin validating this proposed method, we initially set up an automated search to retrieve papers from biomedical databases and assessed whether small LLMs could accurately screen these papers using nuanced classification criteria \cite{Dawood2025}. In this paper, we evaluate and compare pre-trained reasoning LLMs’ abilities to analyze and evaluate individual research papers using an extraction template (questionnaire), and benchmark four LLMs' assessments (Gemini 2.5 Pro, Gemini 2.5 Flash, GPT-5, and GPT-5 Nano) with that of a panel of experts versed in biomedical science. One specific potential MCP was used for this validation: Human Mammary Tumor Virus (HMTV) / Mouse Mammary Tumor Virus-Like Virus (MMTV-LV) and breast cancer. Our objectives were to determine whether LLMs could (1) interpret nuanced biomedical language, (2) critically appraise research, (3) apply causal criteria to expert standards, and (4) produce accurate, trustworthy outputs with minimal hallucinations or omissions, and ultimately utilize our findings to highlight opportunities for improvement, with potential applicability across other biomedical research domains.

\section{Methods}
\subsection{Extraction Template Development}
Rigorous evidence syntheses require transparent and reproducible methodology, including details of how studies are appraised and analyzed \cite{Gough2020}. We developed a structured, standardized extraction template (questionnaire) to ensure repeatable and thorough information extraction. The template was designed to: (1) control and increase transparency of LLM outputs by documenting processes and reasoning, (2) improve reasoning via chain-of-thought (CoT) prompting \cite{Wei2022}, (3) ensure consideration of factors beyond each paper’s stated primary results, and (4) enable comparison between human and LLM responses. The template consisted of pre-prompt context describing the LLM's role and six question sections addressing different aspects of the paper, namely \textit{``Relevancy"}, \textit{``Paper Bibliographic Details"}, \textit{``Paper Integrity and Reliability"}, \textit{``Summary of Paper Contents"}, \textit{``Strength of Evidence"}, and \textit{``Microbial Oncogenesis Criteria"} (see Table~\ref{tab:template_sections} for more details on each section). 

The template included different question types, with 22 multiple-choice questions (MCQs), four Likert scales, 10 multi-select questions, and 40 long-answer (free-text) questions. MCQs refer to questions where only one option can be selected, while multi-select questions refer to questions where any number of options can be selected concurrently. Completing all MCQs, Likert scales, and 14 specific long-answer questions was mandatory. Twenty-six long-answer questions and all 10 multi-select questions were conditional, based on the response to the previous question. Questions were iteratively refined prior to final experiments from team discussions and pilot sessions. The full extraction template can be found in the supplementary material.

\begin{table}[htb]
    \centering
\caption{Overview of the contents of each question section in the extraction template.}
\label{tab:template_sections}
\begin{tabular}{p{0.2\linewidth}p{0.7\linewidth}}
\toprule
\multicolumn{2}{c}{\textbf{Extraction template question sections and their contents}}\\
\midrule
\textbf{Question section} & \textbf{Contents} \\
\midrule
\textbf{Relevancy} &  Relevancy classification criteria followed by questions asking for a classification of the paper (MCQ) and an explanation of why (long-answer).\\
\midrule
\textbf{Paper \mbox{Bibliographic} Details} & Questions for details to trace papers, including the title, publication date, journal of publication, study design, country of origin, and DOI.\\
\midrule
\textbf{Paper Integrity and Reliability}& Questions to determine the reliability of the paper, including whether references were appropriate, and if there was any evidence of a conflict of interest, data or image manipulation, or any contradictions, with a final overall appraisal via a paper reliability score (Likert scale) from 1-5.\\
\midrule
\textbf{Summary of Paper Contents} & Questions for more details of the paper, including the aim and hypothesis, sample characteristics, whether the study's findings are applicable to African populations, and important methodological factors of the study.\\
\midrule
\textbf{Strength of \mbox{Evidence}} & Questions to determine the strength of the paper's evidence, including how data was collected, the appropriateness of the sample size, whether statistical analysis was appropriate, if there were any errors in analysis or presentation of results, and whether there are any limitations to using the paper's findings as evidence for or against the MCP, with a final question asking for an overall appraisal via a strength of evidence score (Likert scale) from 1-5.\\
\midrule
\textbf{Microbial \mbox{Oncogenesis} \mbox{Criteria}} & Descriptions of each criterion followed by questions on whether the paper fulfilled, refuted, left the criterion uncertain, or did not examine the criterion, as well as why (e.g., ``Appropriate methodology/tools", ``Insufficient power"), and to extract the evidence for or against the criterion from the paper. The values for each of the criterions (1 for fulfilled, -1 for refuted, and 0 for either not investigated or left uncertain) are summed to produce the cumulative oncogenesis score (0-10, treated as a Likert scale), with a final question asking for a rating of the paper's impact on the MCP's plausibility (Likert scale) from 1-5\\
\bottomrule
\end{tabular}
\end{table}

\subsection{Human-Validated Test Dataset Creation}
\subsubsection{Locating Papers}
The MCP of HMTV/MMTV-LV and breast cancer was chosen for validation of the system for four reasons. First, its carcinogenicity in humans is unconfirmed, which reflects the intended use-case for our pipeline. Second, there is conflicting evidence for this MCP, enabling us to see how the LLMs would interpret and rate positive, negative, and neutral findings. Third, limited evidence for this MCP in the African context, which is a common challenge across many MCPs under investigation, provides an opportunity to evaluate how the system interprets and applies findings from foreign studies to Africa. Finally, the combination of human and animal research (mouse models) for this MCP allowed us to determine whether LLMs could interpret and apply findings from animal models to humans. 

Twenty-four original research papers that could assist in determining whether HMTV/MMTV-LV causes breast cancer in humans, with differing methodologies and the type and strength of evidence provided, were selected from the PubMed database \cite{FawadKhalid2021, Naccarato2019, Pereira2020, Tabriz2013, Shariatpanahi2017, Morales-Snchez2013,Kulkarni2013,Cedro-Tanda2014,Reza2015,Naushad2017,Perzova2017,AlDossary2018,Lessi2020,Gupta2021,Stewart2022,Gupta2022,Khalid2023,James2024,Callahan2012,Goedert2006,Indik2007, Mazzanti2015,Otten1988,Kincaid2018}. These 24 papers were drawn from the classification dataset (\textit{Relevant} category) \cite{Dawood2025}, consisting of an initial 20 that very clearly fulfilled the \textit{Relevant} classification criteria, supplemented by four papers selected for their unique and nuanced study designs (i.e., where a \textit{Relevant} article may initially seem unrelated to the topic, and require deeper biomedical knowledge to discern its potential effect on the MCP's plausibility), enabling evaluation of LLM performance under a variety of conditions.

\subsubsection{Expert Recruitment}
The extraction template consists of two categories of analytical tasks: (1) questions requiring extensive domain-specific knowledge, and (2) extractive, rule-based, or statistical appraisal questions requiring formal research training but not domain specialization. Category one questions were answered by domain experts, with three experts per paper from a group of seven experts (RD, NI, NAI, KN, JN, EEN, and RP), all of whom were recruited from the University of the Witwatersrand and included specialist clinicians and scientists in oncology, clinical microbiology, pathology, infectious diseases, and virology. Category two questions, which were objective and/or extractive in nature and therefore only required one evaluator, were answered by a clinician-scientist trainee with training in data analytics and biostatistics (KK). These included extraction of bibliographic details of the paper (e.g., title, publication date, publishing journal), identifying contradictions (further evaluated during team discussions to ensure agreement), and bullet-point summaries of the study samples (see extraction template in supplementary material for questions in each category).

\subsubsection{Expert Response Collection}
Author EKS created a web application for the purposes of this study (available at \url{data.oncovectra.com}) that served as a united interface for extraction and appraisal, with dual display of the extraction template questions and the research paper. Papers were uploaded to the platform and assigned to experts based on training specialty and research interest suitability, while ensuring three experts per paper. Experts received a video tutorial on how to use the platform. Experts were aware that the study aimed to compare LLM and domain expert responses, but were blinded to LLM outputs during annotation. The human-validated test dataset is available in the supplementary material.

\subsection{LLM Selection}
We chose four LLMs across two leading LLM vendors: Google (Gemini 2.5 Pro \cite{Comanici2025} and Gemini 2.5 Flash \cite{Comanici2025}) and OpenAI (GPT-5 \cite{singh2026openaigpt5card} and GPT-5 Nano \cite{singh2026openaigpt5card}). This included two frontier reasoning models (Gemini 2.5 Pro and GPT-5) selected for their competitive performance on various reasoning and medical benchmarks at the time of model selection and two lightweight and much cheaper models (Gemini 2.5 Flash and GPT-5 Nano). All four selected LLMs had context-windows $\geq$128K to allow for input of the research paper and extraction template, and had multi-modal vision capability to facilitate interpretation of images and graphical data.  Default parameters were used for all models, except for those related to thinking or reasoning effort, where the highest level was used. Initial outputs (using full versions of all 24 papers and the complete extraction template) were generated on 17 November 2025, using stable Gemini API versions gemini-2.5-pro and gemini-2.5-flash, and GPT-5 model API snapshots gpt-5-2025-08-07 and gpt-5-nano-2025-08-07. All subsequent experiments were done using the same APIs.

\subsection{Generation of LLM Outputs}
LLMs evaluated the 24 papers separately, receiving the full extraction template and one of the 24 research papers in each API call. Text outputs were converted to JSON for analysis. The extraction template instructed models to only answer the extraction template if the paper was relevant to the research question (based on provided relevancy criteria). In cases where LLMs did not classify the paper as relevant, it was rerun with a modified prompt excluding this line to enable assessment of the LLM's answers to the remaining questions.

\subsection{Assessing LLM Outputs}
\label{sec:assessing_llm_outputs}
Krippendorff’s Alpha was initially used to measure inter-rater agreement, however, since this was calculated across three (inter-expert) or four (expert-LLM) raters per question instance, differences in a single response had a large impact on agreement, but did not reflect the broader trends in the data. For example, despite experts reaching consensus (either two or all three experts agreeing) in 92.26\% (310/336) of MCQs (Fig.~\ref{fig:exp_agree_level}), Krippendorff's Alpha for MCQ questions was 0.34, indicating poor expert agreement \cite{Marzi2024}. Therefore, question-specific metrics were designed. For clarity, question instances refer to individual questions and their responses (e.g., question 19 for paper one), while question items refer to the question as found in the extraction template (e.g., question 19 across all papers).

To determine whether LLMs performed at an expert level, their impact on the inter-expert agreement score distribution was assessed. If LLMs responded to the extraction template at the level of domain experts (i.e., provided responses similar to that of experts), then integrating their responses into the inter-expert group produced a distribution that was either indistinguishable from or higher than the inter-expert distribution. If LLM responses did not resemble that of experts (i.e., behaved as an outlier), then integration of their responses would produce a distribution significantly lower than that of the inter-expert agreement distribution. LLM rater groups were created by including the specified LLM's response with the expert responses, and are referred to by LLM name or collectively as the expert-LLM groups.

\subsubsection{Multiple-Choice Questions (MCQ)}
For each MCQ, all inter-expert, expert-LLM, and inter-LLM response pairs were compared, producing a score of one if choices matched and zero if they did not. These scores were then averaged to produce a normalized score per question instance for each rater group, referred to as the \textbf{\textit{MCQ match proportion}}. Additionally, frequency of question instances with full expert agreement (three matching responses), partial agreement (two matching responses), and no agreement (no matching responses) was determined.

\subsubsection{Likert-Scale Questions} 
Considering response pairs could differ by a single rating level and still have good agreement; match proportions were deemed inappropriate. Likert-scale questions were therefore assessed in two ways:
\begin{enumerate}
    \item To determine the distance between response values, for each question instance the absolute differences of response pairs (e.g., $| LLM_xScore - Expert_yScore |$) were averaged for each rater group to produce a normalized difference, referred to as the \textbf{\textit{scale distance}}. 
    \item The scale distance is interpreted differently from other question type metrics, with lower values indicating greater agreement. Therefore to determine cumulative performance across all question instances from all question types, scale distances were inverted to enable assimilation (e.g., if the scale question was out of 5, and the scale distance was 1, the scale score would be 5 - 1 = 4). This is referred to as the \textbf{\textit{inverted distance}}.
\end{enumerate}

Considering the four Likert-scale question items were high impact (providing scores for paper reliability, strength of evidence, cumulative oncogenesis causal criteria score, and the impact of the paper on the MCP) sub-analysis was done for each question item. 

\subsubsection{Multi-select Questions}
All multi-select questions were in the \textit{``Microbial Oncogenesis Criteria"} template section, providing potential explanations (e.g., ``Insufficient power", ``Concerns regarding paper reliability", and ``Appropriate methodology/tools") for why a specific microbial oncogenesis criterion was supported, refuted, or remained uncertain. However, preliminary assessment of expert match rates revealed systematic expert disagreement (Fig.~\ref{fig:exp_agree_level}), with experts selecting opposing options (e.g., ``Appropriate methodology/tools" and ``Inappropriate methodology/tools"). LLMs often returned either all provided options without indicating selection (Gemini Flash 2.5 and GPT-5-Nano), or provided a written explanation for each option. Therefore, no further analysis was performed, as these findings indicate confusion surrounding the question type (for both experts and LLMs) that precludes LLM assessment.

\subsubsection{Long-Answer Questions}
Only question instances with three valid expert responses were used, with 75 question instances across eight question items assessed quantitatively and qualitatively (Table~\ref{tab:question_reference_numbers}).

\begin{table}[htb]
    \centering
    \begin{threeparttable}
\caption{Long-answer question items included in LLM assessment and their question numbers.}
\label{tab:question_reference_numbers}
\begin{tabular}{p{0.09\linewidth}p{0.81\linewidth}}
\toprule
\multicolumn{2}{c}{\textbf{Long-answer question items with their corresponding question numbers}}\\
\midrule
Question Number &  Question\\
\midrule
1.1\tnote{*}& Briefly explain why the paper is relevant, somewhat relevant, or irrelevant.\\
\midrule
 18&List the most important factor(s) in the methodology that influence(s) the strength of the study’s findings.\\\midrule
 26.1\tnote{**}&If there are limitations, what are they?\\\midrule
 28.1.4&State which kind of evidence is provided according to the tools to consider categories and identify the specific finding/s that fulfil or refute this criterion. (\textit{Epidemiologic Association})\\\midrule
 28.2.4&State which kind of evidence is provided according to the tools to consider categories and identify the specific finding/s that fulfil or refute this criterion. (\textit{Histopathologic Association})\\\midrule
 28.4.4&Identify the specific finding/s that fulfil or refute this criterion.  (\textit{Experimental evidence of facilitation of oncogenesis})\\\midrule
 28.5.4&State which kind of evidence is provided according to the tools to consider categories and identify the specific finding/s that fulfil or refute this criterion. (\textit{Molecular and Multi-omics evidence for interaction})\\\midrule
 28.9.4&Identify the specific finding/s that fulfil or refute this criterion.  (\textit{Plausibility})\\ \bottomrule
 \end{tabular}    
 \begin{tablenotes}
    \item [*] Q1.1 follows Q1, which includes the detailed classification criteria.
     \item[**] Q26.1 follows Q26: \textit{``Are there any limitations to using this research as evidence for or against a plausible causative relationship between HMTV/MMTV-like virus and breast cancer?''}
 \end{tablenotes}
\end{threeparttable}
\end{table}

\paragraph{Quantitative Analysis of Long-Answers}
Long-answer responses were scored in two ways using an LLM judge (GPT-5, high-reasoning effort) to improve efficiency, supported by evidence that such models can approximate human judgment when assessing LLM outputs \cite{Croxford2025, Reese2026}. GPT-5 was selected due to its strong benchmark performance, and its scoring was validated on a subset of responses from each LLM to ensure reliability (see supplementary material Table~\ref{tab:1v1_human_score_validation_results} and Table~\ref{tab:3v1_human_score_validation_results}). In both scoring methods, the LLM judge was blinded to the model being assessed when LLM responses were included. The potential impact of self-preference bias (whereby LLM judges favor their own responses) was considered negligible, as the LLM judge was instructed to score the overlap of response pairs rather than identify the superior response. The scoring methods were as follows:
\begin{enumerate}
    \item All response pairs were compared and their overlap scored on a continuous scale from 0-4 up to 1 decimal place (see supplementary material for scoring rubric, Fig.~\ref{fig:1v1_assessment_prompt}). These scores were averaged for each rater group to produce a normalized score for each question instance, referred to as the \textbf{\textit{pairwise overlap score}}.
    \item  Each LLM's response was compared to a clearly indicated reference answer (created by combining all three expert responses) and scored on a discrete scale from 0-4 (see supplementary material for scoring rubric, Fig.~\ref{fig:3v1_assessment_prompt}). This score is referred to as the \textbf{\textit{combined overlap score}}.
\end{enumerate}

\paragraph{Qualitative Analysis of Long-Answers}
Manual evaluation was performed to classify gaps between LLM responses and the combined expert reference answer to distinguish between three possibilities: (1) LLM omissions, (2) LLM hallucinations and/or distortions, and (3) paper-supported additions by the LLM that were omitted from the reference answer. Omissions were defined as information included in the expert response(s) that was absent from the LLM response. Hallucinations referred to instances where LLMs provided information not resembling the original research paper (fabrications), while distortions referred to misinterpretations of the text. Paper-supported additions were defined as information included in the LLM response that was absent from the expert responses but supported by the original paper. Hallucinations, distortions, and paper-supported additions were distinguished by confirming whether the additional information was present in the original research paper.

\subsection{Data Analysis}
\label{sec:data_analysis}
Specialized metrics (described in Sec.~\ref{sec:assessing_llm_outputs}) were used to determine inter-rater agreement. Data was dependent and determined to be non-normal using Shapiro-Wilk tests; hence findings were reported as median (IQR) when applicable, with the exception of aggregate performance, where findings were reported as mean$\pm$SD. Per-paper scores were calculated by summing the agreement metric scores (MCQ match proportions, Likert-scale inverted distances, and long-answer pairwise overlap scores) per paper (isolated to question items with 3 expert responses). Aggregate performance was determined by averaging these per-paper scores. Friedman tests were used to determine whether there were differences in the distributions between groups (reported as test statistic $F_r$ and $p$-value), followed by post-hoc two-sided Wilcoxon Signed-Rank test with Bonferroni correction (reported as test statistic $W$ and $p$-value) and calculation of the standardized effect size ($r$) when the group distributional difference (Friedman test) was determined to be significant ($p < 0.05$). Silverman's rule was used to calculate bandwidth for kernel density estimation (KDE) plots, and distributions were extrapolated. A cross-classified mixed-effect model was fitted to determine whether LLMs tended to match specific expert response patterns, with normalized individual pairwise score as the outcome (score range of 0-1, individual pairs e.g., LLM$_1$ to expert$_1$, using MCQ match score, inverted Likert-scale distance, and long-answer pairwise overlap score), with LLM identity, expert identity, and their interaction as fixed effects, with question type as a covariate, and with question and paper numbers as random effects.
All data was analyzed using Python 3.14.

\section{Results}

\subsection{Inter-Expert Agreement}
\subsubsection{MCQ Agreement}
\begin{figure}[h]
    \centering
        \caption{Expert agreement levels for different question types. Expert consensus (consisting of full and partial agreement) was reached in the majority of MCQ and Likert-scale question instances, while most multi-select question instances had no agreement. Overall, experts had full or partial agreement for the majority (406/488, 83.2\%) of question instances (see ``Combined"). The frequency of question instances where all three expert responses matched (``Full agreement"), where two expert responses matched (``Partial agreement"), and where no expert responses matched (``No agreement"), were determined for each question type.}
    \label{fig:exp_agree_level}
    \includegraphics[width=0.5\linewidth]{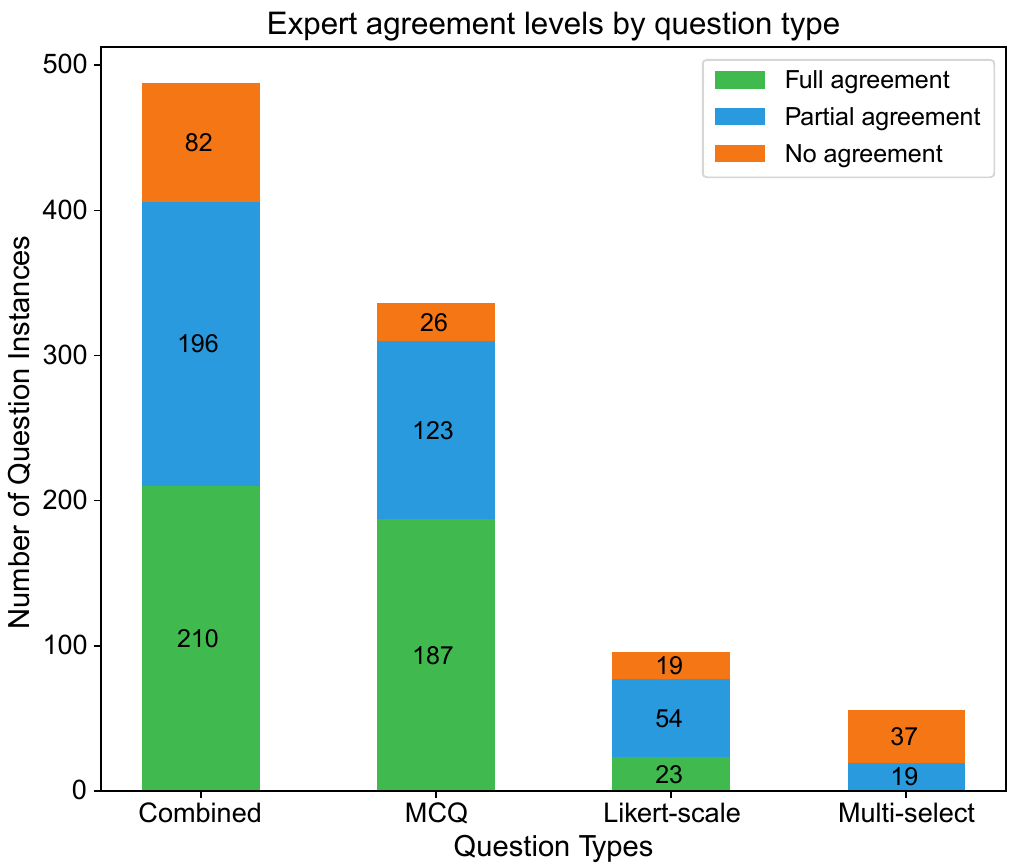}
\end{figure}

Experts displayed strong agreement across MCQs (Fig.~\ref{fig:exp_agree_level}), with full agreement in the majority of question instances (median MCQ match proportion of 1.00, IQR 0.33 - 1.00).  No agreement (full expert disagreement) occurred in 7.7\% (26/336) of MCQ instances and was confined to question items on applicability to African populations (5/336), sample size adequacy (6/336), and five microbial oncogenesis criteria (15/336). Most disagreements in the microbial oncogenesis criteria reflected differences in interpretation rather than opposing conclusions, while 1.2\% (4/336) of instances involved directly contradictory assessments (e.g., one expert concluding that the findings supported microbial carcinogenicity and another concluding that they refuted it).

\subsubsection{Likert-Scale Agreement}
\begin{figure}[!htb]
    \centering
        \caption{Distribution of Likert-scale distances across rater groups. Gemini 2.5 Pro's Likert-scale ratings deviated substantially from experts, resulting in significantly increased scale distances when integrated, while Gemini 2.5 Flash, GPT-5, and GPT-5 Nano provided similar ratings to experts and did not alter the inter-expert scale distance distribution. Likert-scale distances were calculated per question instance (n=96) by averaging the differences between Likert-scale ratings for each rater pair (e.g., expert one and Gemini 2.5 Pro) in the rater group (e.g., Gemini 2.5 Pro). Outliers (circular outlines) were determined using Tukey's method (values outside of interval $Q_1- 1.5 \times IQR$ and $Q_3 + 1.5 \times IQR$), whiskers indicate maximum and minimum values within interval. A Friedman test with post-hoc two-sided Wilcoxon-Signed-Rank tests were used to determine whether distributions differed.}
    \label{fig:Likert-scale_distances}
    \includegraphics[width=0.5\linewidth]{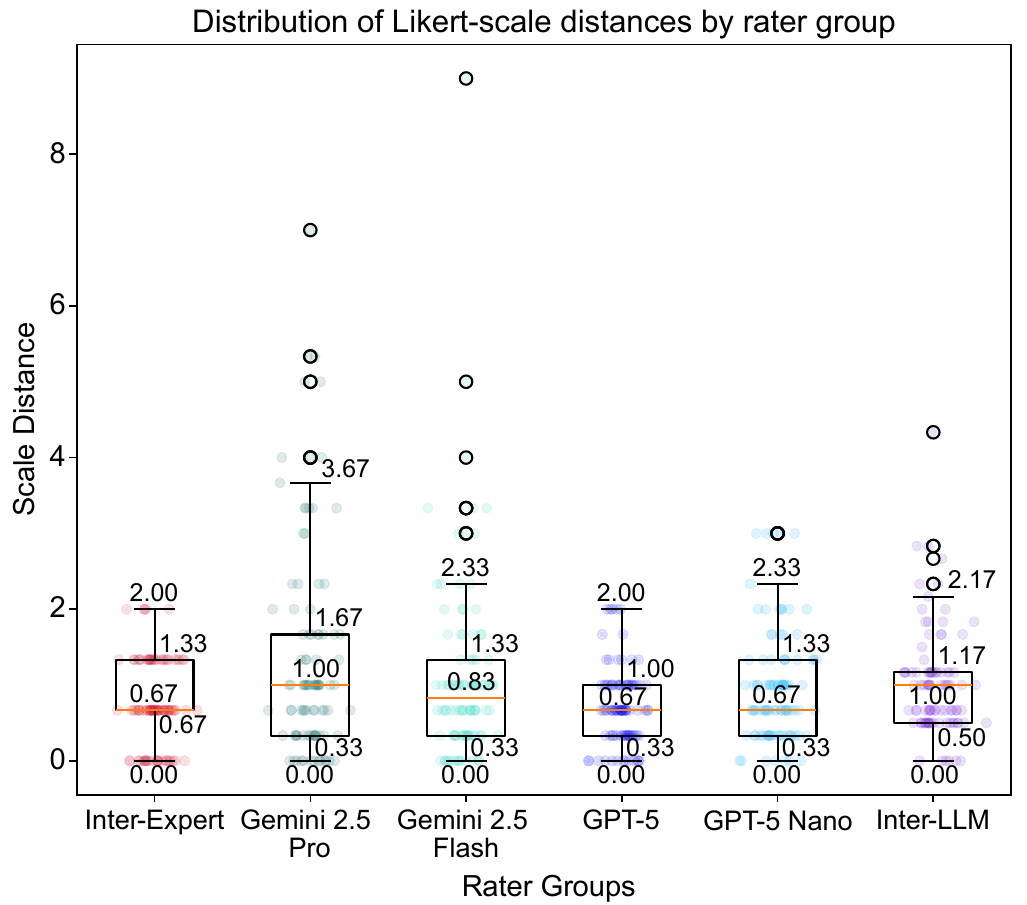}
\end{figure}

Overall, expert Likert-scale ratings were closely clustered. For most question instances, experts had partial agreement with the outlier rating being one level away from consensus (Fig.~\ref{fig:Likert-scale_distances}). Considering that individuals may differ in their interpretation and application of Likert-scale ratings, deviations of one level indicate good agreement around a central value. No agreement (full expert disagreement) occurred in 19.8\% (19/96) of question instances and was most common for cumulative oncogenesis score (7/96) and paper reliability score (5/96). In most instances (15.6\%, 15/96), expert responses spanned three consecutive Likert-scale levels (e.g., 1, 2, and 3), while 4.2\% (4/96) of instances contained an outlier rating two levels removed from the nearest score.

Of the four Likert-scale question items, the strength of evidence and paper impact on MCP plausibility scores had the strongest expert agreement, followed by the paper reliability score, with score clustering and small scale distances. Cumulative oncogenesis score had the most disagreement with larger scale distances, reflecting moderate variability in interpretation and application of microbial oncogenesis criteria.

\subsubsection{Long-Answer Question Agreement}
\begin{figure}[!htb]
    \centering
        \caption{LLMs had higher combined overlap scores (B) than pairwise overlap scores (A), indicating that LLMs tended to include information from each of the experts rather than a single expert. \textbf{(A)} LLM long-answer pairwise overlap score distributions were indistinguishable from the inter-expert distribution, indicating that LLM responses were as similar to individual experts as experts were to each other. Inter-LLM distribution differed significantly from inter-expert and expert-LLM distributions (all $p < 0.0001, \ r > 0.75$), indicating that LLM responses had greater overlap with each other than with experts, and had greater overlap than experts had with each other. Long-answer response pairs for 75 question instances were scored on their overlap by an LLM-judge (GPT-5, blinded to LLM being assessed) on a continuous rubric scale from 0-4. Scores for each question instance were averaged for each rater group to produce the pairwise overlap score. \textbf{(B)} LLM long-answer responses captured most key points found across three expert responses, with LLM responses predominantly scoring $\geq$2. A Friedman test found no significant differences in combined overlap scores across the different LLMs ($F_r = 6.48, \ p =0.0906$), indicating that response completeness did not differ by model size. For 75 question instances, LLM responses were compared to a reference answer created by combining three separate expert responses. Each LLM response was scored on whether it covered the key and/or minor points in the reference answer by an LLM-judge (GPT-5, blinded to LLM being assessed) using a discrete rubric scale from 0-4 (see supplementary material Fig.~\ref{fig:3v1_assessment_prompt} for scoring rubric).}
    \label{fig:long_answer_scoring}
    \includegraphics[width=1\linewidth]{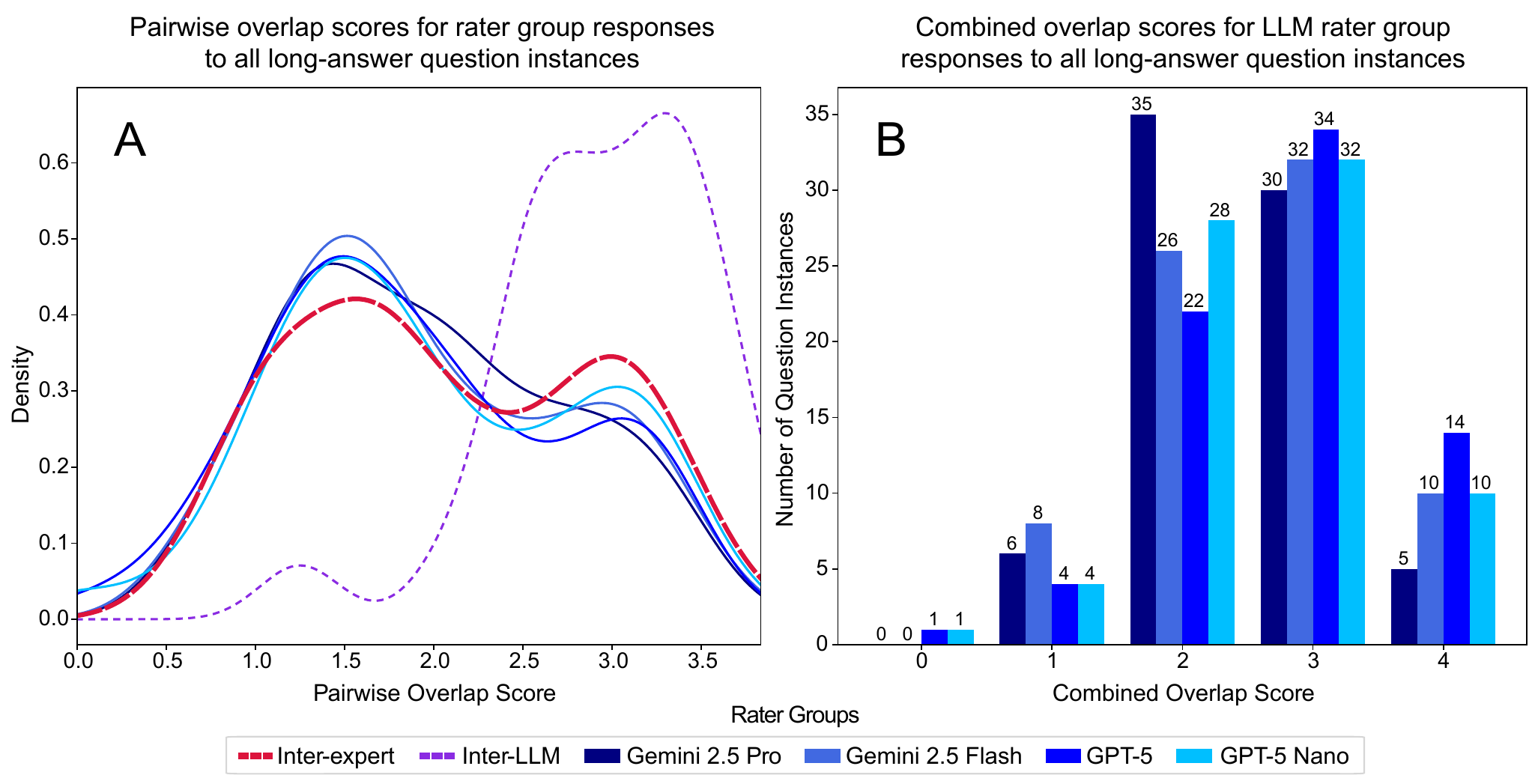}
\end{figure}

Expert responses captured similar key points, but differed on important nuances (Fig.~\ref{fig:long_answer_scoring}A). Pairwise overlap scores differed between question items. Q1.1 (see Table~\ref{tab:question_reference_numbers} for full question items) had the highest scores, indicating strong agreement for reasoning on paper relevancy, while Q18 and Q26.1, which both focused on methodology, had the lowest pairwise overlap scores, indicating minor overlap with significant differences in responses. Manual evaluation found that experts included non-contradicting points focusing on different aspects of the paper, thereby producing a comprehensive combined response. Therefore, assessment of LLM responses using the combined overlap score method would mitigate these differences. Question items 28.1.4, 28.2.4, 28.4.4, 28.5.4, and 28.9.4, were combined for analysis due to their small sample size (six combined) and conceptual similarity (involved applying and extracting evidence for or against the microbial oncogenesis criteria). Manual evaluation of responses for these question items determined differences to be due to varying response foci and detail levels. Similarly to Q18 and Q26.1, this indicates that experts focused on different aspects of the question item and the paper's findings, collectively producing a well-rounded response.

High levels of consensus and complementary responses among experts indicated that question items were clear and strengthened confidence in the human-validated test dataset as a means of benchmarking the LLMs.

\subsection{Overall LLM Performance Across Extraction Template}
\subsubsection{Aggregate Performance}
\begin{figure}[!htb]
    \centering
        \caption{Average per-paper scores across rater groups. GPT-5 and GPT-5 Nano responses integrated seamlessly with expert responses across the 24 papers, producing no appreciable change in inter-expert agreement, while Gemini 2.5 Pro and Gemini 2.5 Flash's responses were more inconsistent and tended to deviate from that of experts, significantly decreasing inter-expert agreement. Individual per-paper scores were determined by summing the MCQ match proportions, Likert-scale inverted distances, and long-answer pairwise overlap scores for each paper. Individual per-paper scores were averaged to produce the average per-paper score per rater group. Error bars represent $\pm SD$.}
    \label{fig:average_paper_performance}
    \includegraphics[width=0.5\linewidth]{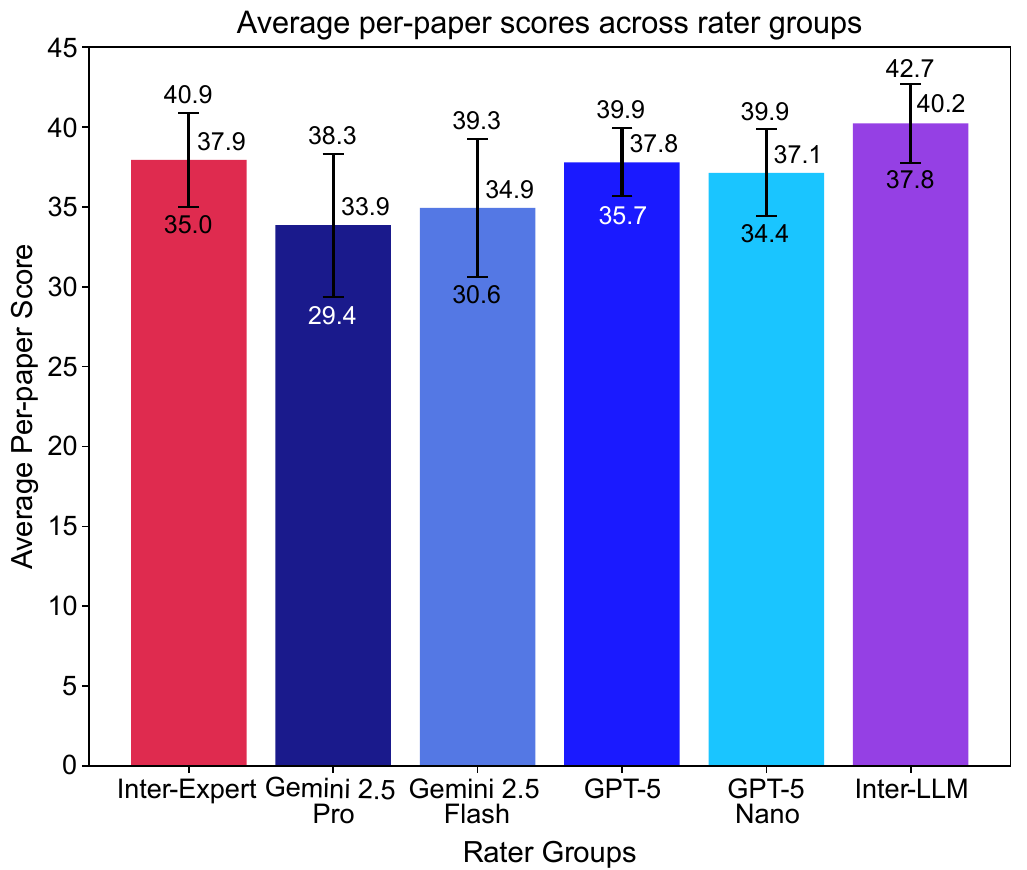}
\end{figure}

Across the 24 papers, GPT-5 ($W = 139.5,\ p =1.0$) and GPT-5 Nano ($W = 115.0,\ p =1.0$) consistently gave similar responses to experts, producing no overall change in inter-expert agreement when integrated (i.e., when LLM responses were included as one of the human experts for analysis of agreement). In contrast, the responses of Gemini 2.5 Pro ($W = 17.0,\ p =0.0004,\ r = 0.78$) and Gemini 2.5 Flash ($W = 39.0,\ p =0.0127, \ r = 0.65$) tended to deviate from experts, resulting in significantly decreased inter-expert agreement when integrated (Fig.~\ref{fig:average_paper_performance}). Inter-LLM agreement was significantly higher than inter-expert agreement ($W = 45.0,\ p = 0.0267, \ r = 0.61$), suggesting that LLM responses were more similar to each other than to expert responses, and exhibited greater inter-response similarity than that observed among experts.

Overall, GPT-5 and GPT-5 Nano produced responses most similar to domain experts, resulting in indistinguishable score distributions, while Gemini 2.5 Pro and Gemini 2.5 Flash tended to produce responses that differed from domain experts, resulting in significantly lower score distributions.

\subsubsection{Performance by Question Type}

\begin{figure}[!htb]
    \centering
        \caption{Distribution of MCQ match proportions across rater groups. All LLMs provided MCQ answers similar to experts, with integration of their responses resulting in either no change in inter-expert agreement (Gemini 2.5 Pro, Gemini 2.5 Flash), or a significant increase in inter-expert agreement (GPT-5, GPT-5 Nano). MCQ match proportions were calculated per question instance (n=336) by averaging the binary match rates for each rater pair (e.g., expert one to Gemini 2.5 Pro) in the rater group (e.g., Gemini 2.5 Pro rater group). Friedman test with post-hoc two-sided Wilcoxon-Signed-Rank tests were used to determine whether distributions differed.}
        \label{fig:MCQ_match_proportions}
    \includegraphics[width=0.5\linewidth]{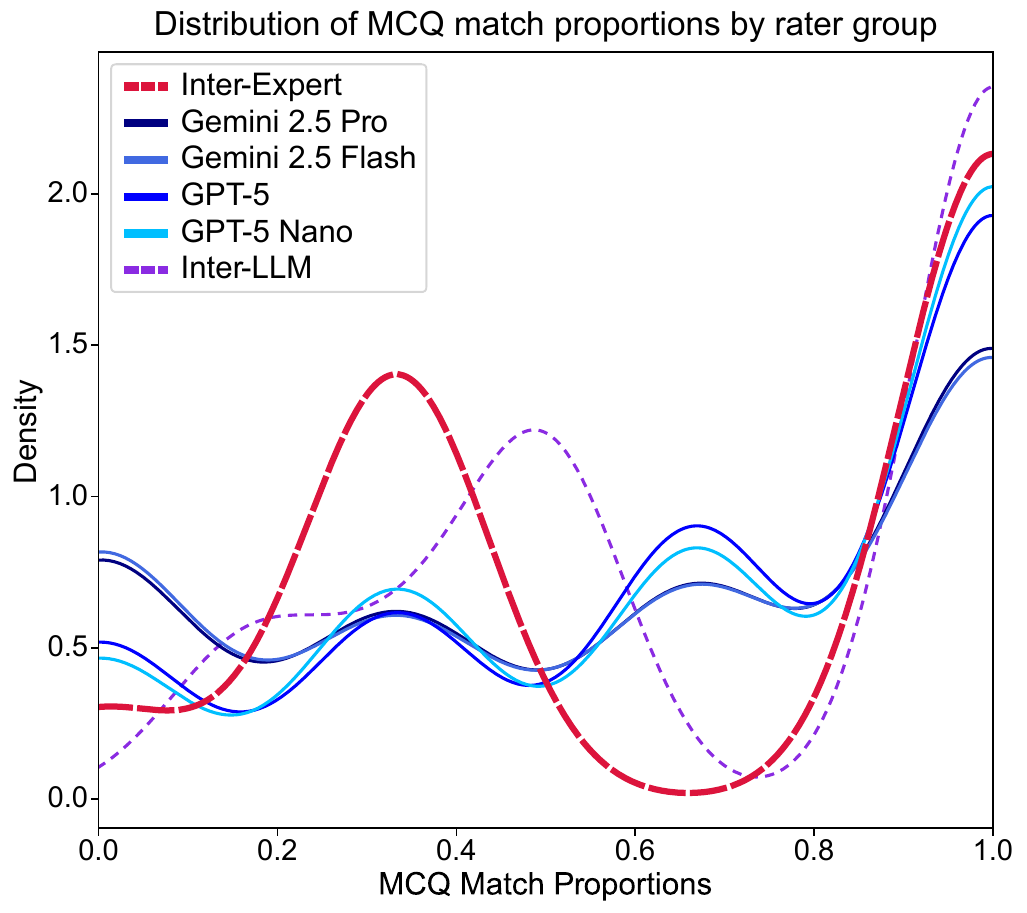}
\end{figure}

\paragraph{MCQ Performance}
All models provided similar MCQ answers to experts. GPT-5 ($W = 3345.0$, $p =0.0106$, $r = 0.76$) and GPT-5 Nano's ($W = 2601.5, \ p =0.0005, \ r = 0.79$) MCQ answers were consistently similar to experts, tending to agree with expert consensus (either partial or full), therefore significantly increasing expert agreement when integrated. Gemini 2.5 Pro ($W = 6110.0, \ p =1.0$) and Gemini 2.5 Flash ($W = 6007.0, \ p =1.0$) tended to agree with the expert outlier, with integration of their MCQ answers having no effect on expert agreement (Fig.~\ref{fig:MCQ_match_proportions}).

\paragraph{Likert-Scale Performance}
Across the four Likert-scale questions, GPT-5 ($W = 929.0, \ p =1.00$) and GPT-5 Nano ($W = 1127.5, \ p =1.00$) rated the papers most similarly to experts, with integration of their ratings producing no overall change in expert scale distances. Gemini 2.5 Pro gave considerably higher ratings than experts, resulting in significantly greater expert scale distances ($W = 639.0, \ p =0.0013, \ r = 0.63$). Similarly, while integration of Gemini 2.5 Flash's ratings did not significantly alter expert scale distances ($W = 769.0, \ p =0.1981$), these ratings were higher than experts', resulting in larger scale distances (Fig.~\ref{fig:Likert-scale_distances}). For all question items, higher ratings equated to greater leniency (for example, greater paper reliability or strength of evidence), indicating a consistent positivity bias in Gemini 2.5 Pro's evaluations, with Gemini 2.5 Flash behaving similarly. Gemini 2.5 Pro and Gemini 2.5 Flash tended to agree with the expert outlier, while GPT-5 and GPT-5 Nano agreed with the expert consensus and outlier with similar frequency.

Overall, GPT-5, GPT-5 Nano, and Gemini 2.5 Flash rated papers similarly to experts across the different Likert-scale question items, although Gemini 2.5 Flash had a tendency to rate papers slightly more positively than experts, while Gemini 2.5 Pro consistently rated papers more favorably than experts, indicating a positivity bias in its evaluations.

\paragraph{Long-Answer Performance}

\subparagraph{Quantitative Performance} 

\textbf{Pairwise Overlap}
LLM responses were as similar to individual expert responses as individual experts were to each other, with indistinguishable pairwise overlap score distributions between expert-LLM and inter-expert rater groups (Fig.~\ref{fig:long_answer_scoring}A, see supplementary material Fig.~\ref{fig:1v1_assessment_prompt} for scoring rubric). LLMs gave very similar responses to one another, with significantly greater overlap between their responses compared to that of the expert-LLM and inter-expert groups (all pairwise tests $p < 0.0001, \ r > 0.75$).

\textbf{Combined Expert Response Overlap}
LLMs produced comprehensive responses that tended to capture key ideas dispersed across the three expert responses rather than aligning with a single expert, achieving greater combined overlap scores than pairwise overlap scores (Fig.~\ref{fig:long_answer_scoring}B, see supplementary material Fig.~\ref{fig:3v1_assessment_prompt} for scoring rubric). Response completeness did not differ by model complexity ($F_r = 6.48, \ p =0.0906$), with smaller models (Gemini 2.5 Flash, GPT-5 Nano) capturing the same amount of key points as larger models (Gemini 2.5 Pro, GPT-5) across all question instances.

Overall, LLM long-answer responses were comprehensive, capturing key ideas from across the three expert responses. LLM responses were very similar, capturing the same amount of key points regardless of model size.

\subparagraph{Qualitative Analysis}
\begin{table}[!htb]
\centering
\begin{threeparttable}
\caption{Of question instances where LLMs included information not in expert responses, almost all were paper-supported additions, with a few hallucinations and/or distortions found in smaller models (Gemini 2.5 Flash, GPT-5 Nano). While omissions were numerous, there were similar counts of paper-supported additions. Across 75 question instances, information gaps between experts and LLMs were manually assessed and categorized as either omissions, hallucinations and/or distortions, or paper-supported additions.}
\label{tab:hallucinations_omissions_accurate}
\begin{tabular}{l >{\centering\arraybackslash}p{2cm} >{\centering\arraybackslash}p{3cm} >{\centering\arraybackslash}p{3cm} >{\centering\arraybackslash}p{2cm}}
\toprule
\multicolumn{5}{c}{\textbf{Frequency of omissions, hallucinations and/or distortions, and paper-supported}} \\
\multicolumn{5}{c}{\textbf{additions in LLM long-answer responses}} \\
\midrule
\textbf{LLM} &   \textbf{Omissions} \tnote{1}
& \textbf{Information not in expert responses}&\textbf{Hallucinations and/or distortions} \tnote{2} & \textbf{Paper-supported additions} \tnote{3} \\
\midrule
Gemini 2.5 Pro &   68 
&51&0 & 51 \\
\midrule
Gemini 2.5 Flash &   64 
&64&2 & 62 \\
\midrule
GPT-5 &   67 
&62&0 & 62 \\
\midrule
GPT-5 Nano &   66 &69&7 & 62 \\
\bottomrule
\end{tabular}
\begin{tablenotes}
    \item[1] \textbf{Omissions} refer to information LLMs excluded from their responses that experts included in theirs.
    \item[2] \textbf{Hallucinations and/or distortions} refer to information included in the LLM response that was inaccurate to the original paper, either fabricated or representing a misinterpretation of the text.
    \item[3] \textbf{Paper-supported additions} refer to information included in the LLM response, absent from expert responses, but faithful to or found within the original paper.
\end{tablenotes}
\end{threeparttable}
\end{table}

\textbf{Omissions}
\label{sec:omissions}
Gemini 2.5 Flash and GPT-5 Nano had slightly less omissions than frontier LLMs Gemini 2.5 Pro and GPT-5 (Table \ref{tab:hallucinations_omissions_accurate}). Per-question item analysis found that omissions for relevancy explanations (Q1.1) and responses involving extraction of findings supporting or refuting the microbial oncogenesis criteria (28.1.4, 28.2.4, 28.4.4, 28.5.4, 28.9.4) were generally minor and reflected differences in response focus between experts and LLMs, with singular exceptions for GPT-5 and GPT-5 Nano for the microbial oncogenesis criteria, while omissions for methodology-related question items (Q18 and Q26.1) were more significant.

For methodology-related question items, LLMs tended to omit the same information. In some instances, experts included irrelevant information (for example, a description of results rather than methodological strengths or weaknesses), however, legitimate omissions included: (1) failure to comment on sample size or appropriateness, and (2) no mention of the country the study was conducted in, although models did occasionally mention the limited geographic generalizability of results. For Q26.1 specifically, LLMs were less likely to comment on potential errors or missing information in the papers than experts. 

Out of all omissions, the most severe occurred in the \textit{``Microbial Oncogenesis Criteria"} section for questions 28.2.4 (for \textit{``Histopathologic Association"}) and 28.4.4 (for \textit{``Experimental Evidence of Facilitation of Oncogenesis"}), by GPT-5 Nano and GPT-5 respectively. For both question instances, all three experts agreed that the paper helped fulfilled these criteria, while these LLMs did not. While both models' reasoning were valid and indicated stricter adherence to the criterion's definition than experts, GPT-5 Nano's reasoning was inconsistent with its responses for this criterion in other papers, although these question instances were not included in the main analysis as they lacked three expert responses for comparison.

Overall, smaller models had less omissions than larger models, with Gemini 2.5 Flash having the least and Gemini 2.5 Pro having the most, although models tended to omit the same information. For most question items omissions were minor, with the exception of methodological questions where LLMs often failed to consider factors surrounding sample size, country of study origin, and potential errors in the papers.

\textbf{Hallucinations and Distortions}
\label{sec:hallucinations}
Frontier models (Gemini 2.5 Pro, GPT-5) had no hallucinations or distortions across the 75 question instances, while smaller models (Gemini 2.5 Flash, GPT-5 Nano) exhibited a limited number of such errors. Of the instances where Gemini 2.5 Flash (64) and GPT-5 Nano (69) introduced information not present in expert responses, 2 and 7 instances were classified as hallucinations and/or distortions (Table \ref{tab:hallucinations_omissions_accurate}). These represented misinterpretations of nuanced, domain-specific methodological details rather than traditional fabrications, and occurred exclusively in methodology-focused questions (Q18 and Q26.1, see Table~\ref{tab:question_reference_numbers} for full questions). None of the errors overlapped between the two models, indicating distinct failure modes. 

\textbf{Gemini 2.5 Flash}
Both errors were classified as distortions with minimal impact on evidence appraisal, and consisted of a misinterpretation of a PCR quality-control step and incomplete pooling of tissue sample type counts. 

\textbf{GPT-5 Nano}
\label{sec:GPT-5 Nano_hallucinations}
GPT-5 Nano repeated errors across the two methodology-related question items for three papers, resulting in four unique hallucinations/distortions out of the seven instances. GPT-5 Nano appeared more prone to text misinterpretation than Gemini 2.5 Flash, with its errors impacting evidence strength appraisal. These included a misunderstanding of standard PCR terminology, a misreading of a 2$\times$2 contingency table, conflation of metastatic status with overall cancer prevalence, and falsely stating that appropriate controls were absent, which GPT-5 Nano characterized as a critical methodological flaw weakening reliability.

Overall, frontier models Gemini 2.5 Pro and GPT-5 had no hallucinations and/or distortions in the 75 question instances assessed, while smaller models Gemini 2.5 Flash and GPT-5 Nano had minimal instances. Both smaller models' errors represented misinterpretations of domain specific methodologies, however GPT-5 Nano's had slightly more errors which were more severe, with downstream effects on the evaluation of the paper's strength of evidence.

\textbf{Paper-Supported Additions}
\label{sec:paper_supported_additions}
LLMs tended to include the same paper-supported additions as one another, even for methodology-related question items (Q18 and Q26.1) requiring applied reasoning over extraction. The majority of question instances for these question items had paper-supported additions. For Q18, LLMs tended to extract more detailed methodological information (e.g., the exact controls and processes used), while for Q26.1, LLMs tended to include additional limitations which required integration of paper-based information and domain knowledge (e.g., acknowledging that FFPE tissues have inherent nucleic acid degradation risks that may impact viral detection). Although LLMs tended to include similar limitations to one another, often repeating the same (yet contextually appropriate) set of limitations, there were instances where the LLMs noted pivotal points that the experts missed.

\subsection{Performance by Template Section}

\subsubsection{Relevancy}
Gemini 2.5 Pro was maximally sensitive, classifying all 24 papers as \textit{Relevant}. Similarly, Gemini 2.5 Flash classified one paper as \textit{Somewhat Relevant} (aligned with expert consensus) and GPT-5 classified two papers as \textit{Somewhat Relevant}, one aligning with consensus and one where experts agreed the paper was \textit{Irrelevant}. GPT-5 Nano demonstrated the strictest classification threshold, labeling 54.17\% (13/24) of papers as \textit{Somewhat Relevant}. In most cases, this aligned with at least a single expert; however, in one instance all experts classified the paper as \textit{Relevant}, and in another all agreed it was \textit{Irrelevant}. 

LLM relevancy explanations consistently captured all major expert-identified points, with minor omissions (pairwise overlap scores $\geq$2 and combined overlap scores $\geq$3). Qualitative analysis of GPT-5 Nano's responses found that it consistently identified relevant mechanistic or associative findings in the papers, explaining the high pairwise overlap scores, but concluded that absence of temporality or direct causal proof warranted a \textit{Somewhat Relevant} classification.

Overall, LLMs demonstrated understanding of domain-specific nuances (both in the paper and in the classification criteria) and strong alignment with expert reasoning on paper relevancy.

\subsubsection{Paper Integrity and Reliability}

\paragraph{Identifying Irrelevant References}
Only GPT-5 claimed to find irrelevant references, and this was for two papers. Manual re-evaluation found GPT-5 to be incorrect for the first paper, although this seemingly had no downstream effect on GPT-5's evidence appraisal. However, for the second paper, GPT-5 identified three references that did appear to be irrelevant to the sections they were cited in, which initial human evaluation missed.

\paragraph{Identifying Potential Conflicts of Interest}
GPT-5 Nano claimed that university funding constituted ``a potential COI concern in some assessments", while Gemini 2.5 Flash claimed that the private laboratories the authors were affiliated with were contracted with the National Cancer Institute (NCI), which ``could introduce an indirect, potential commercial interest". In many instances, LLMs - most notably GPT-5 Nano - cited the paper's own no conflict of interest declaration as reasoning for their response. 

\paragraph{Identifying Contradictions}
\label{sec:contradictions}
LLMs had difficulty consistently detecting human-identified contradictions in full-text papers, although these had minimal impact on the paper's reliability. Gemini 2.5 Flash and GPT-5 were the only models to identify any human-identified contradictions, and this was for one of the 13 instances. Gemini 2.5 Pro and GPT-5 identified three and five contradictions, respectively, that the human evaluator missed. However, Gemini 2.5 Pro, GPT-5, and GPT-5 Nano had one, two, and four instances, respectively, where they either fabricated or misinterpreted a contradiction. 

Gemini 2.5 Pro's hallucination stated that values for a subgroup analysis did not add up to the correct value, yet in its response it demonstrated that they did, thereby fabricating a contradiction, while GPT-5 missed the authors' justification for methodological decisions. GPT-5 Nano's hallucinations were more severe, and included false inconsistencies in result values, sample sizes, and the counts in a $2\times 2$ contingency table.

To determine whether failure arose from context window limitations or true inability, models were given the isolated sections of text, figures, and/or tables in which the human-identified contradictions were found. After repeated attempts, Gemini 2.5 Pro did not return any results, citing API rate limits despite request limit adherence. This isolation method improved performance considerably, with Gemini 2.5 Flash (11/13), GPT-5 (12/13), and GPT-5 Nano (9/13) detecting the majority of human-identified contradictions, indicating that prior failure was likely due to context window limitations rather than lack of ability. One of these contradictions required interpretation of an electrophoresis gel image, which Gemini 2.5 Flash and GPT-5 demonstrated, despite being generalist models without biomedical specialization. One contradiction, where the paper was inconsistent in which mouse models were used, was missed by all LLMs while the rest of the missed contractions varied between Gemini 2.5 Flash and GPT-5 Nano. In two of these isolated sections, Gemini 2.5 Flash and GPT-5 each identified two additional contradictions (a total of four each) that were missed by the human evaluator, with a total of six additional unique contradictions identified.

Overall, LLMs were largely unable to find the human-identified contradictions within the full-text papers, although Gemini 2.5 Flash and GPT-5 did identify contradictions the human evaluator missed. Gemini 2.5 Pro, GPT-5, and GPT-5 Nano all hallucinated some contradictions, although GPT-5 Nano's were the only severe ones. Isolation experiments improved performance considerably, indicating that prior failure was due to context window limitations rather than inability or lack of domain knowledge.

\begin{figure}[!htb]
    \centering
        \caption{Distribution of paper reliability scores given by different rater groups (Likert-scale ratings, n=24). All three expert ratings for each question instance have been aggregated into the ``Experts" group. Stacked bars represent the percentage of question instances assigned each score (1-5) by each rater group. Bars are aligned at 0\% by the central rating value (3, dotted line), with lower scores (representing lower paper reliability) extending to the left, and higher scores (representing higher paper reliability) extending to the right. Gemini 2.5 Pro and Gemini 2.5 Flash gave very similar paper reliability scores to experts (predominantly 5s), while GPT-5 and GPT-5 Nano in particular gave lower paper reliability scores than experts (predominantly 3s and 4s), indicating more strict critique of paper reliability (although not significantly). Likert-scale rating meanings have been truncated for visualization (see supplementary material for extraction template with full meanings).}
    \label{fig:reliability_score}
    \includegraphics[width=0.5\linewidth]{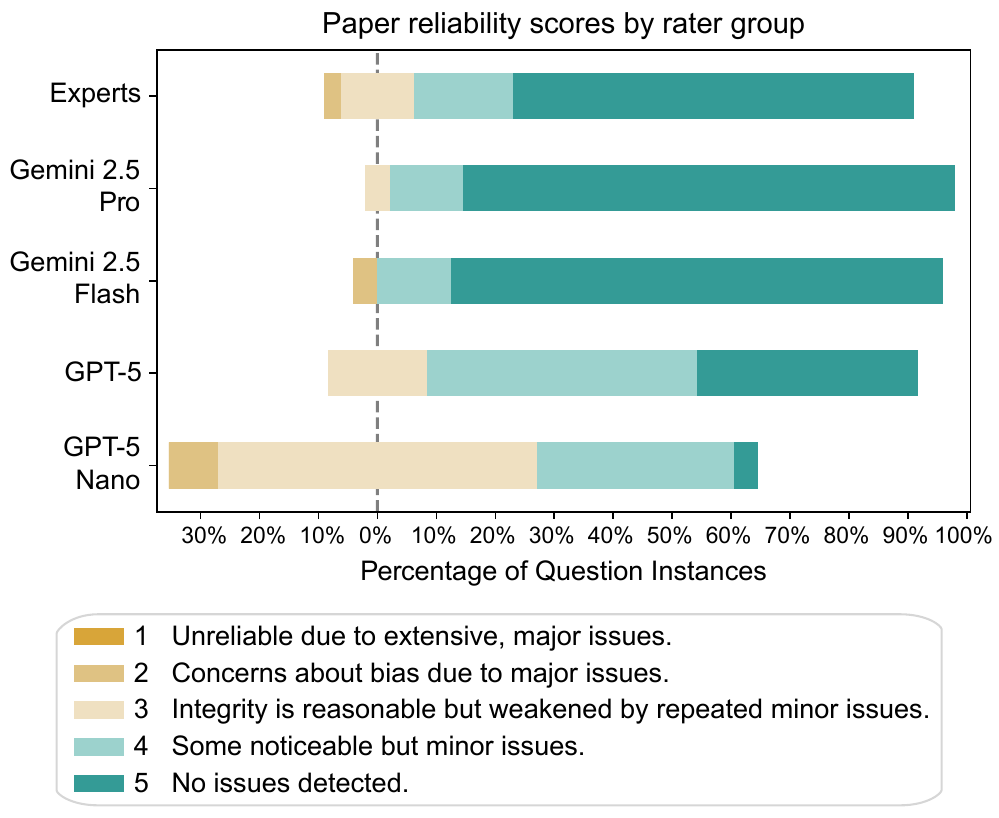}
\end{figure}

\paragraph{Paper Reliability Score}
All models rated paper reliability similarly to experts, with no significant differences in the scale distances between the inter-expert and expert-LLM group pairs. Gemini 2.5 Pro and Gemini 2.5 Flash demonstrated the closest alignment with expert paper reliability ratings, followed by GPT-5, while GPT-5 Nano showed the greatest divergence, tending to rate paper reliability slightly lower than experts (Fig.~\ref{fig:reliability_score}).

\subsubsection{Summary of Paper Contents}
\begin{figure}[!htb]
    \centering
        \caption{LLMs had higher combined overlap scores (B) than pairwise overlap scores (A) for Q18, indicating that LLMs tended to include important methodological factors from multiple experts rather than a single expert. Inter-LLM pairwise overlap scores were considerably greater than other rater groups (A), indicating that LLMs often included the same methodological factors as one another in their responses. \textbf{(A)} LLM long-answer pairwise overlap score distributions were near indistinguishable from the inter-expert distribution. Long-answer response pairs for 75 question instances were scored on their overlap by an LLM-judge (GPT-5, blinded to LLM being assessed) on a continuous rubric scale from 0-4 (see supplementary material, Fig.~\ref{fig:1v1_assessment_prompt} for scoring rubric). Scores for each question instance were averaged for each rater group to produce the pairwise overlap score. \textbf{(B)} LLM long-answer responses captured most key points found across three expert responses, with LLM responses predominantly scoring $\geq$2. For 75 question instances, LLM responses were compared to a reference answer created by combining three separate expert responses. Each LLM response was scored on whether it covered the key and/or minor points in the reference answer by an LLM-judge (GPT-5, blinded to LLM being assessed) using a discrete rubric scale from 0 to 4 (see supplementary material Fig.~\ref{fig:3v1_assessment_prompt} for rubric).}
    \label{fig:methodological_factors}
    \includegraphics[width=1\linewidth]{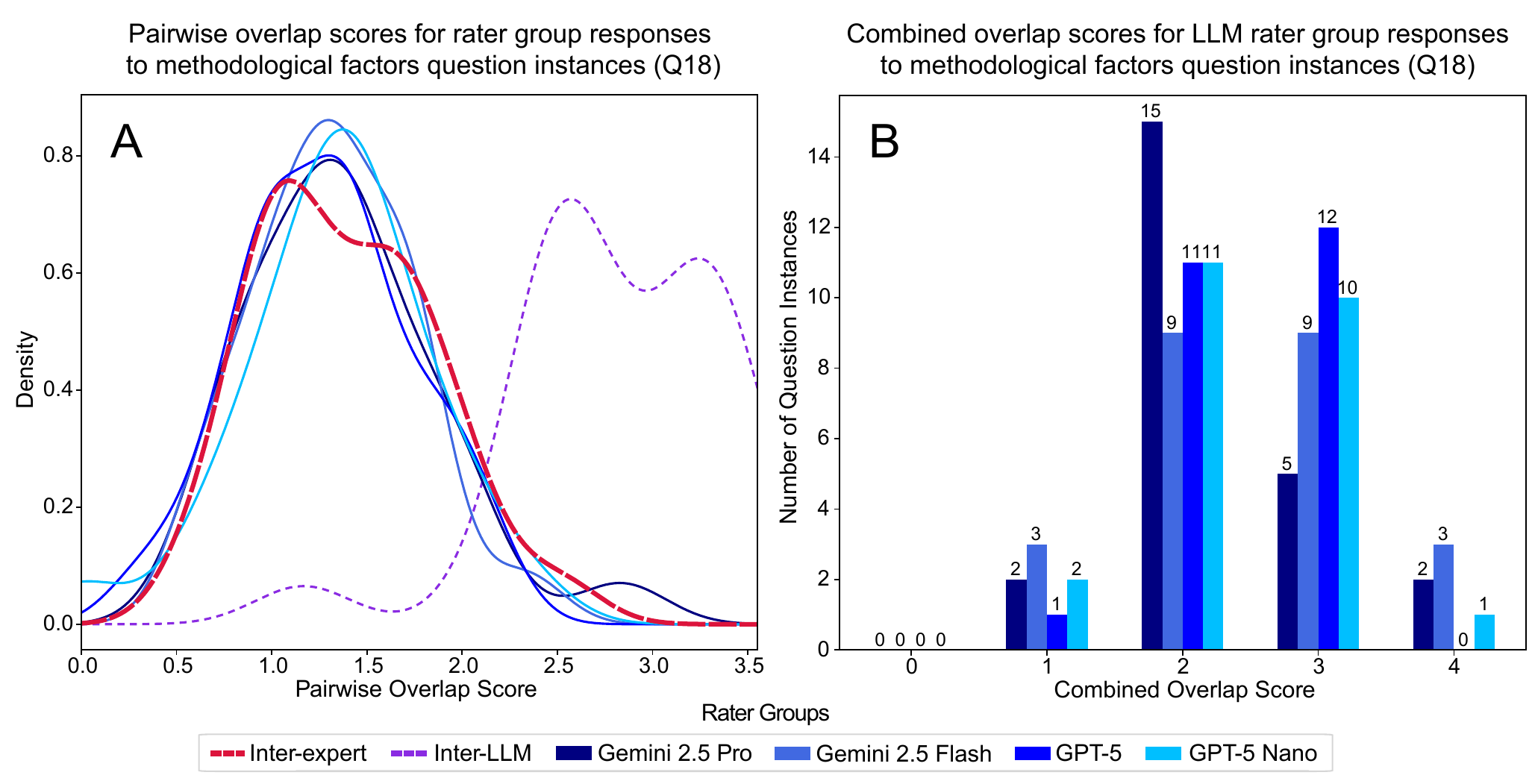}
\end{figure}

\paragraph{Identifying Important Methodological Factors Influencing the Strength of Findings}
Expert responses for this question item tended to include different, non-contradicting focus points, resulting in low pairwise overlap scores for inter-expert and expert-LLM groups (Fig.~\ref{fig:methodological_factors}A). However, LLMs tended to integrate key methodological considerations from multiple experts rather than aligning exclusively with a single expert, achieving higher combined overlap scores than pairwise overlap scores (Fig.~\ref{fig:methodological_factors}B). Gemini 2.5 Flash and GPT-5 demonstrated slightly better agreement with the combined reference response compared to Gemini 2.5 Pro and GPT-5 Nano, although distributions were broadly similar across models (Fig.~\ref{fig:methodological_factors}B). The majority of LLM responses had combined overlap scores $\geq$2, indicating capture of some key points with omissions (see Sec.~\ref{sec:omissions}). Gemini 2.5 Flash and GPT-5 Nano had hallucinations and/or distortions in some of their responses (see Sec.~\ref{sec:hallucinations}), indicating difficulty interpreting nuanced biomedical research methodology. 

\subsubsection{Strength of Evidence}

\paragraph{Identifying Errors in Statistical Analysis or Result Presentation}
GPT-5 and Gemini 2.5 Flash alone identified one of the four human-identified errors in result presentation (percentage miscalculation), and Gemini 2.5 Pro, GPT-5, and GPT-5 Nano identified two, three, and one errors, respectively, that were missed by the human evaluator. The human-identified errors the LLMs missed included two instances of incorrectly presented Venn diagrams (difference values did not exclude intersection values and intersection values were duplicated), and another instance where a results table had an incorrect category label. Additionally, Gemini 2.5 Pro and GPT-5 Nano both repeated hallucinated and/or distorted errors they gave in their contradiction analysis (see Sec.~\ref{sec:contradictions} for more details). 

\begin{figure}[!htb]
    \centering
        \caption{LLMs had higher combined overlap scores (B) than pairwise overlap scores (A) for Q26.1, indicating that LLMs tended to include limitations from multiple experts rather than a single expert. Higher inter-LLM pairwise overlap scores (A) indicated that LLMs included similar limitations to one another. \textbf{(A)} LLM long-answer pairwise overlap score distributions were similar to the inter-expert distribution. Long-answer response pairs for 75 question instances were scored on their overlap by an LLM-judge (GPT-5, blinded to LLM being assessed) on a continuous rubric scale from 0-4 (see supplementary material Fig.~\ref{fig:1v1_assessment_prompt} for scoring rubric). Scores for each question instance were averaged for each rater group to produce the pairwise overlap score. \textbf{(B)} LLM long-answer responses captured most key points found across three expert responses, with LLM responses predominantly scoring $\geq$2. For 75 question instances, LLM responses were compared to a reference answer created by combining three separate expert responses. Each LLM response was scored on whether it covered the key and/or minor points in the reference answer by an LLM-judge (GPT-5, blinded to LLM being assessed) using a discrete rubric scale from 0 to 4 (see supplementary material Fig.~\ref{fig:3v1_assessment_prompt} for scoring rubric).}
    \label{fig:limitations}
    \includegraphics[width=1\linewidth]{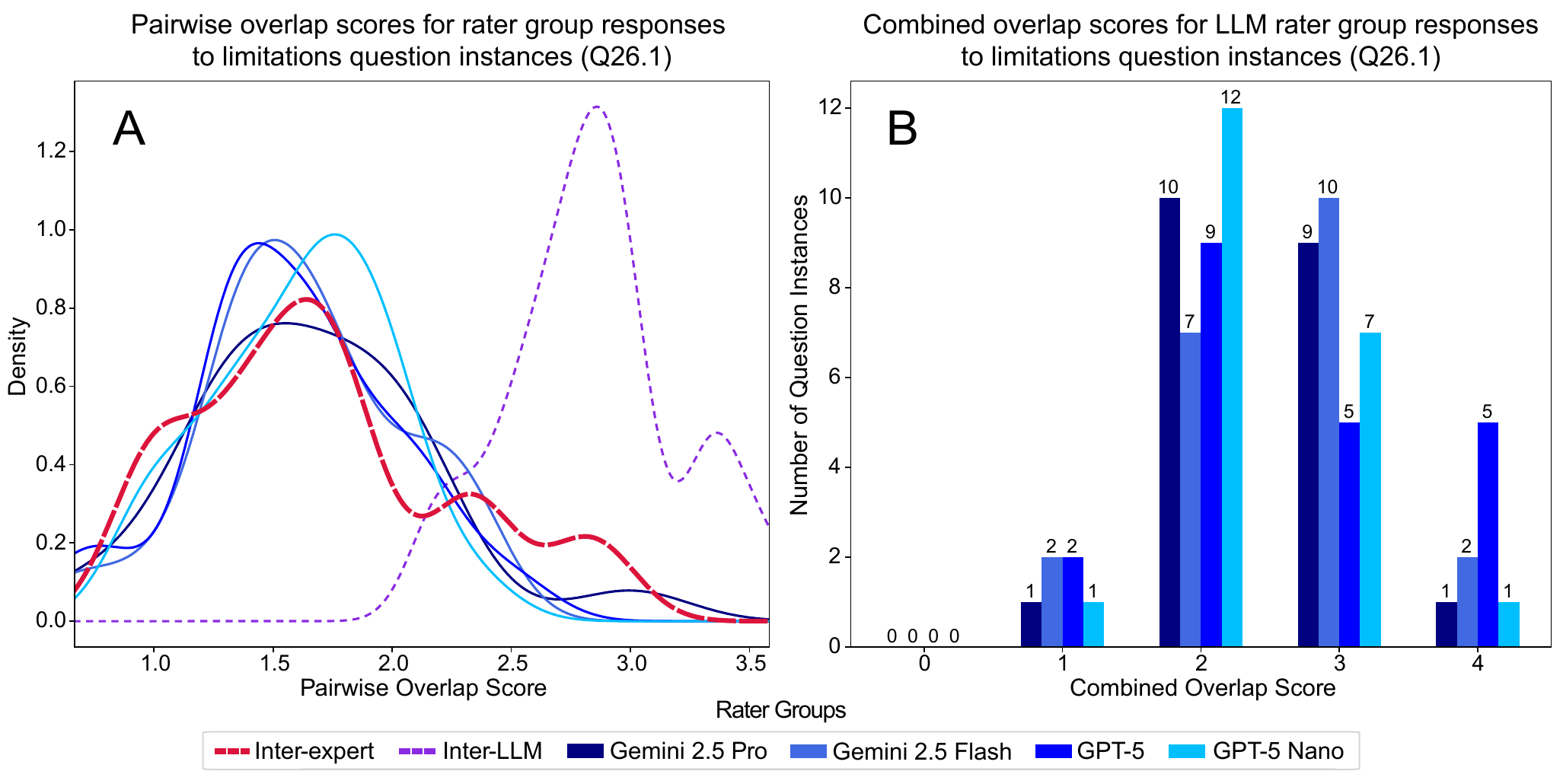}
\end{figure}

\paragraph{Identifying Limitations}
LLMs tended to emphasize similar limitations to one another, while experts generally provided different (yet complementary) limitations, resulting in substantially higher inter-LLM pairwise overlap scores compared to the inter-expert and expert-LLM rater groups (Fig.~\ref{fig:limitations}A). LLMs tended to integrate limitations from multiple experts rather than aligning with a single expert, with higher combined overlap scores than pairwise overlap scores for all LLMs (Fig.~\ref{fig:limitations}B).

\begin{figure}[!htb]
    \centering
        \caption{Distribution of strength of evidence scores given by different rater groups (Likert-scale ratings, n=24). All three expert ratings for each question instance have been aggregated into the ``Experts" group. Stacked bars represent the percentage of question instances assigned each score (1-5) by each rater group. Bars are aligned at 0\% by the central rating value (3, dotted line), with lower scores (representing lower strength of evidence) extending to the left, and higher scores (representing higher strength of evidence) extending to the right. Gemini 2.5 Pro and Gemini 2.5 Flash tended to rate strength of evidence slightly higher than experts (predominantly 3s and above), indicating greater leniency, while GPT-5 tended to rate strength of evidence slightly lower (predominantly 2s and below), indicating more strict evidence appraisal.}
    \label{fig:strength_of_evidence_score}
    \includegraphics[width=0.5\linewidth]{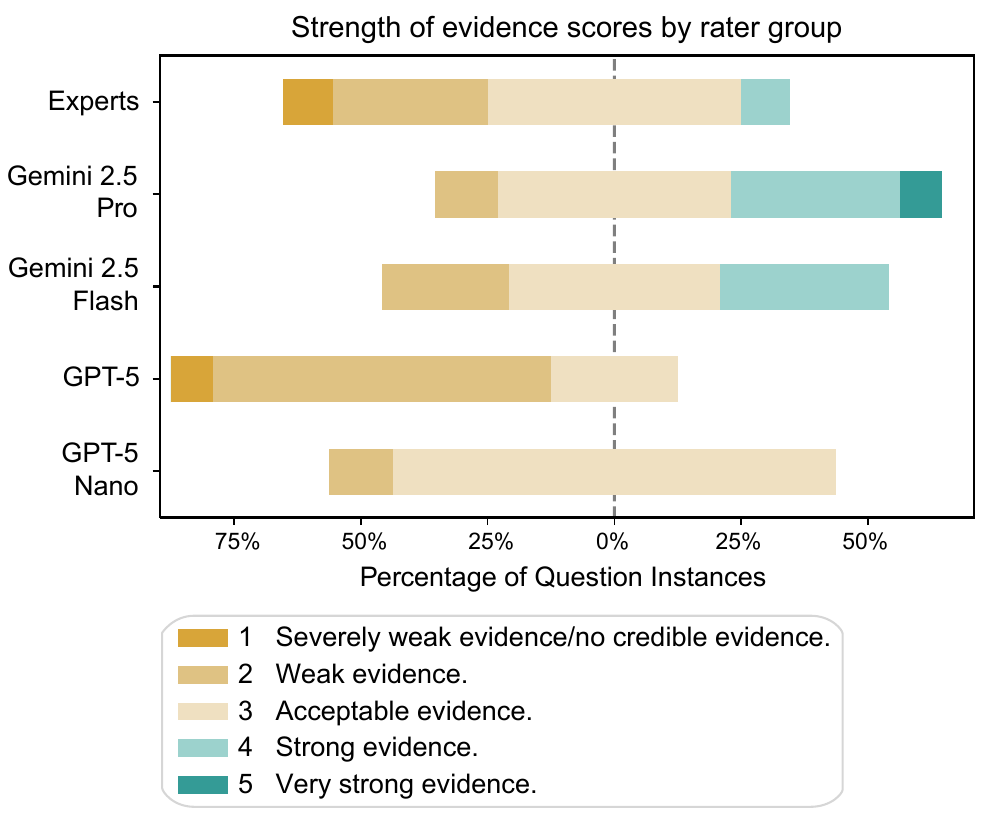}
\end{figure}

\paragraph{Strength of Evidence Score}
All models rated strength of evidence similarly to experts. Most LLM ratings were the same as or one level away from expert consensus (all LLM median scale distances $\leq$1), with no significant differences in the scale distances across the rater groups ($F_r = 13.42$, $p =0.0197$). When ratings did deviate from experts, Gemini 2.5 Pro and Gemini 2.5 Flash tended to rate strength of evidence slightly higher than experts, indicating greater leniency, while GPT-5 tended to rate strength of evidence slightly lower than experts, indicating a slightly stricter interpretation of evidence strength (Fig.~\ref{fig:strength_of_evidence_score}).

\subsubsection{Microbial Oncogenesis Criteria}

\paragraph{Identifying Evidence Type and Findings Supporting/Refuting Microbial Oncogenesis Criteria}
Across the six questions analyzed, LLMs tended to include details from all three expert responses,  with higher combined overlap scores than pairwise overlap scores. GPT-5 had the highest combined overlap scores (3.00, IQR 1.50 - 3.00), followed by Gemini 2.5 Pro (2.50, IQR 2.00 - 3.00) and Gemini 2.5 Flash (2.00, IQR 2.00 - 2.75), with GPT-5 Nano having the lowest scores (2.00, IQR 1.25 - 2.00), indicating that it tended to miss key points. Despite generally high performance, GPT-5, along with GPT-5 Nano, each had an instance where, contrary to expert consensus, they did not perceive a paper as helping to fulfill or refute the specified criterion, possibly indicating that these LLMs were occasionally more strict than experts in applying microbial oncogenesis criteria. Neither Gemini 2.5 Pro nor Gemini 2.5 Flash had this issue. 

\begin{figure}[!htb]
    \centering
        \caption{\textbf{(A)} Distribution of cumulative oncogenesis scores given by different raters (Likert-scale ratings, n=24). All three expert ratings for each question instance have been aggregated into the ``Experts" group. Stacked bars represent the percentage of question instances assigned each score (0-10) by each rater group. Bars are aligned at 0\% by the central rating value (5, dotted line), with lower scores extending to the left, and higher scores extending to the right. Cumulative oncogenesis scores were determined by summing scores for each of the 10 microbial oncogenesis criteria. Higher scores approaching 10 indicated fulfillment of more microbial oncogenesis criteria. Lower scores approaching 1 indicated either: (1) fulfillment of less criteria, or (2) fulfillment of some criteria combined with evidence to refute other criteria, as combinations of positive scores for certain criteria (paper helps support specific criterion) and negative scores for other criteria (paper helps refute specific criterion) may cancel out to produce a low positive score or a score of 0 depending on the combination. GPT-5 and GPT-5 Nano had cumulative oncogenesis scores most similar to experts (between 0 and 4), while Gemini 2.5 Pro and Gemini 2.5 Flash tended to state that papers supported more microbial oncogenesis criteria (in some instances $\geq$5 criteria for a single paper), and therefore gave higher cumulative oncogenesis scores than experts. \textbf{(B)} Comparison of paper impact on MCP plausibility scores (Likert-scale ratings, n=24) given by different raters. All three expert ratings for each question instance have been aggregated into the ``Experts" group. Stacked bars represent the percentage of question instances assigned each score (1-5) by each rater group. Bars are aligned at 0\% by the central rating value (3, dotted line), with lower scores (representing that the paper more strongly refuted the MCP's plausibility) extending to the left, and higher scores (representing that the paper more strongly supported the MCP's plausibility) extending to the right. Gemini 2.5 Pro, Gemini 2.5 Flash, and GPT-5 all tended to rate the paper's impact on the MCP's plausibility more highly, but Gemini 2.5 Pro and Gemini 2.5 Flash both gave more extreme ratings (1s for ``Strongly refutes" and 5s for ``Strongly supports") compared to experts, GPT-5, and GPT-5 Nano.}
    \label{fig:cumulative_oncogenesis_mcp_impact}
    \includegraphics[width=1\linewidth]{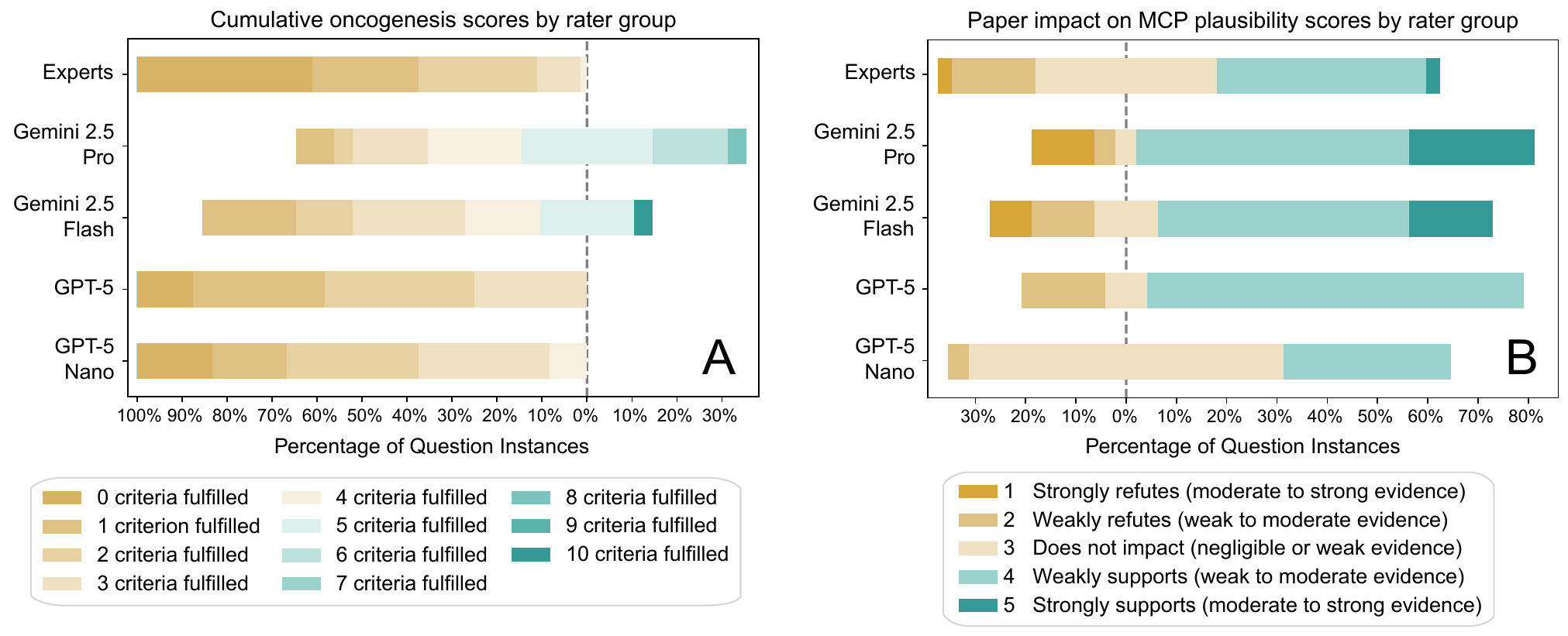}
\end{figure}

\paragraph{Cumulative Oncogenesis Score}
GPT-5 ($W$ $= 86.0$, $\textit{p} = 1.0$) and GPT-5 Nano ($W$ $= 60.0$, $\textit{p} = 1.0$) applied microbial oncogenesis criteria in a manner similar to experts, with inclusion of their cumulative oncogenesis scores leaving inter-expert agreement unchanged. Gemini 2.5 Pro ($W = 3.0$, $p =0.0004$, $r = 0.86$) and Gemini 2.5 Flash ($W = 21.0$, $p =0.0091$, $r = 0.75$) were far more lenient than experts in stating a paper helped fulfill various microbial oncogenesis criteria, giving higher Likert-scale ratings and therefore significantly decreasing expert agreement (Fig.~\ref{fig:cumulative_oncogenesis_mcp_impact}A). 

Gemini 2.5 Pro and Gemini 2.5 Flash had a tendency to state papers helped fulfill a given microbial oncogenicity criterion in contexts where evidentiary support was indirect or absent. For example, for a review of publicly available ecological data correlating mouse population changes with breast cancer incidence \cite{Stewart2022}, both Gemini models assigned cumulative oncogenesis scores of 8/10 and 10/10 respectively, indicating that they considered the paper to satisfy nearly all microbial oncogenesis criteria, including temporality, reproducibility/validation, dose-response relationship, molecular/multi-omics interaction, and even prevention. Notably, this behavior did not reflect misunderstanding of the papers, but rather an expansive interpretation of what constituted fulfillment of the oncogenesis criteria, with scoring justifications suggesting a susceptibility to ecological inference bias and false-cause reasoning.

Overall, GPT-5 and GPT-5 Nano applied the microbial oncogenesis criteria most similar to experts, while Gemini 2.5 Pro and Gemini 2.5 Flash deviated from experts, applying the criteria much more leniently.

\paragraph{Rating Impact of Paper on MCP}
LLMs rated each paper's impact on the MCP's plausibility similarly to experts, with no significant differences in the scale distances across the different rater groups ($F_r = 10.79$, $p =0.0557$), however, Gemini 2.5 Pro, Gemini 2.5 Flash, and GPT-5 all had a tendency towards rating the paper's impact on the MCP's plausibility higher, indicating slightly greater leniency in this regard than experts (Fig.~\ref{fig:cumulative_oncogenesis_mcp_impact}B).

\subsection{LLM-Expert Similarity Pattern Analysis}
A cross-classified mixed effects model investigating the pairwise scores of all question types (see Sec.~\ref{sec:data_analysis}) found no statistically significant evidence that any LLM preferentially matched specific experts (all $p > 0.7$), indicating that LLMs did not replicate the reasoning patterns of specific domain experts.

\subsection{Stability of GPT-5 Outputs}
As GPT-5's responses aligned most closely with experts, repeat experiments were performed to determine output stability. We ran 30 repeats on two papers, using a subset of six MCQs and two Likert-scale question items (see supplementary material Sec. 1.4 for included question items and the dataset for GPT-5's outputs).

GPT-5 demonstrated high response stability across both papers. For the first paper, six of eight question items received identical responses across all runs, resulting in an overall stability of 92.5\% (222/240 responses). For the second paper, five of eight question items received identical responses across all runs, resulting in an overall stability of 92.1\% (221/240 responses). Across both papers, overall response stability was 92.3\% (443/480 responses).

Where variation occurred, GPT-5's responses generally remained closely aligned with expert assessments. For several MCQs, GPT-5 alternated between responses that matched either the expert consensus or the expert outlier. Similarly, variation in Likert-scale ratings was typically limited to a single rating level and remained close to expert evaluations.

Overall, despite occasional deviations, GPT-5’s responses were consistent across repeated runs and generally remained within the range of expert interpretations, indicating stable expert-aligned reasoning across repeated evaluations.

\section{Discussion}
\begin{figure}[!htb]
    \centering
        \caption{Summary of LLM performance by question type, overall, and on the core objectives. \textit{``GPT-5 models"} refers to GPT-5 and GPT-5 Nano. \textit{``Gemini models"} refers to Gemini 2.5 Pro and Gemini 2.5 Flash. \textit{``Larger models"} refers to Gemini 2.5 Pro and GPT-5, while \textit{``Smaller models"} refers to Gemini 2.5 Flash and GPT-5 Nano.}
    \label{fig:summary_of_LLM_performance}
    \includegraphics[width=1\linewidth]{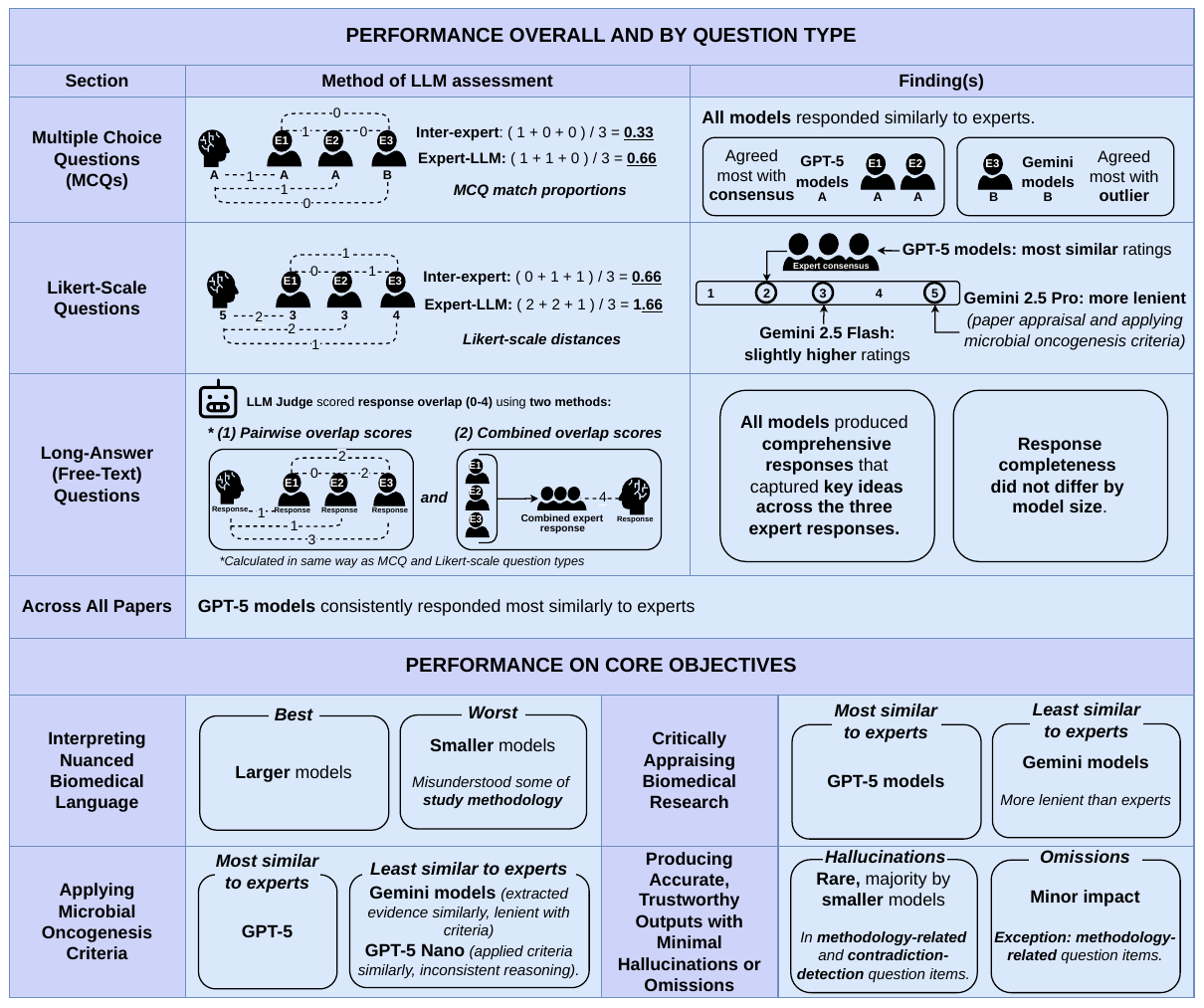}
\end{figure}

Our results show that pre-trained reasoning LLMs are capable of analyzing and evaluating individual biomedical research papers at the level of domain experts on most tasks when using a structured extraction and appraisal template, with minimal hallucinations (see Fig.~\ref{fig:summary_of_LLM_performance} for summary of results).

The developed human-validated test dataset had strong inter-rater reliability, with high rates of full and partial agreement for MCQs ($>90\%$) and Likert-scale questions ($>80\%$, Fig.~\ref{fig:exp_agree_level}), and close clustering of Likert-scale ratings even when consensus was absent (Fig.~\ref{fig:Likert-scale_distances}). Long-answer responses demonstrated moderate overlap (Fig.~\ref{fig:long_answer_scoring}A), with variability primarily reflecting complementary reasoning rather than substantive disagreement between experts, as expected with nuanced domain-specific tasks. Considering this variability, we assessed whether inclusion of LLM responses altered the overall distribution of expert agreement (i.e., whether LLMs behaved as an additional expert), rather than requiring exact matches to expert responses. This may be a more effective strategy for evaluating LLMs in domains characterized by complex, nuanced tasks with non-contradicting inter-expert variability.

The exception to this pattern of strong inter-rater reliability was the multi-select (i.e., select all that apply) question type, which proved challenging for both experts and LLMs. Experts occasionally provided directly conflicting responses (e.g., one selecting ``Appropriate methodology/tools" and another selecting ``Inappropriate methodology/tools"), potentially reflecting differences in domain expertise and interpretation of study strengths and limitations. LLMs also struggled with this format, frequently selecting all available options or generating long-answer responses rather than selections, potentially reflecting poor exposure to this question type in their training data. These question items were excluded from further analysis as expert disagreement precluded the establishment of a reliable reference standard. The fact that both experts and LLMs encountered difficulties suggests that the issue lay primarily with the question type rather than either evaluator group. Future iterations of the extraction template will therefore replace multi-select items with a series of binary (yes/no) statements for each microbial oncogenesis criterion, accompanied by a justification field. Based on these findings, multi-select question formats are not recommended for either human or LLM-based evaluation of biomedical literature.

Across all 24 papers, GPT-5 and GPT-5 Nano performed most consistently and similarly to experts, while Gemini 2.5 Pro ($W = 22.0$, $p < 0.0001$, $r = 0.75$) and Gemini 2.5 Flash ($W = 39.0$, $p =0.0127$, $r = 0.65$) behaved as outliers, significantly decreasing inter-expert agreement (Fig.~\ref{fig:average_paper_performance}). However, performance varied across question types and items, with methodological appraisal questions and identification of contradictions being the most persistent areas of vulnerability. 

Across MCQ items, LLMs performed similarly to experts - consistent with prior evidence that generalist LLMs (particularly GPT models) perform well on biomedical MCQ answering \cite{Siam2025}, with GPT-5 ($W = 3345.0$, $p =0.0106$, $r = 0.76$) and GPT-5 Nano ($W = 2601.5$, $p =0.0005$, $r = 0.79$) having significantly increased inter-expert agreement, agreeing most with expert consensus. Across Likert-scale questions, Gemini 2.5 Pro ($W = 639.0$, $p =0.0013$, $r = 0.63$) had the greatest divergence from experts, significantly increasing scale distances by giving considerably higher Likert-scale ratings, indicating greater leniency in paper evaluation, while Gemini 2.5 Flash, GPT-5, and GPT-5 Nano gave similar scores to experts and therefore did not alter inter-expert distributions (Fig.~\ref{fig:Likert-scale_distances}). When Gemini 2.5 Pro and Gemini 2.5 Flash's ratings did match an expert's, it tended to be the outlier, while GPT-5 and GPT-5 Nano matched the expert consensus and outlier with similar frequency. Repeat evaluations with GPT-5 (30 runs on two papers for a subset of eight question items) demonstrated response consistency across MCQ and Likert-scale question types (overall average response stability of 92.3\%), with responses consistently matching or remaining closely clustered around expert ratings.

LLM long-answer responses were as similar to individual experts as experts were to each other, resulting in LLM long-answer pairwise overlap score distributions being indistinguishable from expert distributions (Fig.~\ref{fig:long_answer_scoring}A). Despite operating at a disadvantage against three domain experts, LLMs captured the majority of key points identified collectively by the expert group, demonstrating greater alignment with the combined expert reference answers than with individual expert responses (Fig.~\ref{fig:long_answer_scoring}B), suggesting integrative synthesis across perspectives and expertise. Despite differences in model size, complexity, and reported reasoning capacity, no significant superiority was observed across models in combined overlap scores, indicating functional convergence in synthesis capacity - however, limited instances of hallucinations and/or distortions by smaller models Gemini 2.5 Flash and GPT-5 Nano reduced their trustworthiness compared to frontier models Gemini 2.5 Pro and GPT-5 (see Sec.~\ref{sec:hallucinations}). 

Interestingly, LLMs exhibited greater agreement with one another than with experts, and greater agreement than was observed between experts themselves. However, this effect was driven entirely by long-answer questions; agreement patterns for MCQ and Likert-scale items were comparable between LLMs and experts. Among the model pairings, Gemini 2.5 Pro and Gemini 2.5 Flash showed the greatest overlap in long-answer responses, followed by Gemini 2.5 Flash and GPT-5, and then GPT-5 and GPT-5 Nano. High inter-LLM agreement may partially reflect similarities in training data or learned representations, although the strong agreement observed between Gemini 2.5 Flash and GPT-5 suggests that shared LLM developer lineage alone cannot fully explain the findings. Another likely contributor is response length: LLMs typically provided substantially more detailed answers than experts, often listing multiple observations where experts highlighted only one or two key points. Consequently, LLM responses had a greater opportunity to overlap with one another, whereas agreement between experts and LLMs was inherently constrained by the concise nature of expert responses.

Given the inclusion of experts from multiple specialties, a cross-classified mixed-effects model was used to determine whether LLMs preferentially aligned with specific expert perspectives, thereby exhibiting specialty-specific reasoning patterns. However, there was no statistically significant evidence that any LLM consistently matched particular experts more closely than others (all $p > 0.7$). This finding suggests that the models did not replicate the reasoning patterns of individual domain specialists. One possible interpretation is that the LLMs integrated perspectives across multiple areas of expertise, producing responses that reflected a synthesis of expert viewpoints rather than alignment with any single specialist. This interpretation is consistent with the long-answer combined overlap scores, where LLMs achieved greater overlap with the combined expert responses than with any individual expert. An alternative explanation is that model responses were relatively random and thus did not systematically resemble any specific expert. However, the consistency of LLM performance across question types, together with their strong performance on long-answer questions, argues against largely random or inconsistent reasoning patterns and lends greater support to the integrative synthesis hypothesis. Nevertheless, the ability to detect expert-specific alignment may have been limited by the study design, as each paper was evaluated by only three experts and expert participation overlapped only partially across the dataset.

Building on the test dataset reliability and overall LLM performance assessment, we return to the core objectives: whether generalist, pre-trained LLMs can
\begin{enumerate}
    \item Interpret nuanced biomedical language,
    \item Critically appraise biomedical research,
    \item Apply microbial oncogenesis causal criteria to expert standards, and
    \item Produce accurate, trustworthy outputs with minimal hallucinations or omissions
\end{enumerate}

\subsection{Interpreting Nuanced Biomedical Language}
While LLMs appeared to generally understand nuanced biomedical language, Gemini 2.5 Flash and GPT-5 Nano occasionally misinterpreted domain-specific methodological terminologies, leading to distortions and - in the case of GPT-5 Nano - downgrading of paper evidence (see Sec.~\ref{sec:GPT-5 Nano_hallucinations}). While smaller generalist models do not appear able to fully interpret nuanced biomedical language to the standard of domain experts, frontier models Gemini 2.5 Pro and GPT-5 performed sufficiently well, demonstrating understanding of complex domain-specific information, strengthening trust in their capabilities and enabling their use for systematic biomedical evidence synthesis processes such as ours. 

\subsection{Critically Appraising Biomedical Research}
Systematic differences in appraisal behavior were observed across models, suggesting architecture-dependent biases. Gemini 2.5 Pro significantly increased scale distances and demonstrated a consistent tendency toward more favorable appraisals of paper reliability (Fig.~\ref{fig:reliability_score}), strength of evidence (Fig.~\ref{fig:strength_of_evidence_score}), and microbial oncogenesis plausibility (Fig.~\ref{fig:cumulative_oncogenesis_mcp_impact}A). Gemini 2.5 Flash behaved similarly, however, it did not produce significant changes in inter-expert agreement. In contrast, GPT-based models exhibited relatively more conservative scoring behavior that predominantly matched that of experts. These findings converge with our prior work demonstrating increased leniency in Gemini models when applying classification criteria \cite{Dawood2025}. While leniency increased sensitivity in the literature screening stage, it presently reduced alignment with expert judgment. In evidence synthesis contexts, such differences could meaningfully influence cumulative grading, prioritization, and downstream decision-making. 

GPT-5 models demonstrated the highest alignment with expert critical appraisal. As all experts were recruited from a single institution, it is possible that some degree of shared institutional perspective influenced expert critical appraisal. However, several factors suggest that this is unlikely to explain the observed agreement. First, the expert panel was not educationally homogeneous, comprising individuals who received their undergraduate and postgraduate training from different institutions and who therefore entered the study with diverse academic backgrounds and methodological perspectives. Second, experts represented multiple specialties and departments, each of which place emphasis on different approaches to evidence evaluation, study design, and interpretation. This diversity was reflected in the fact that there was variance across the expert responses, and that they did not demonstrate complete consensus, with occasional disagreement in their assessments of paper reliability, strength of evidence, and microbial oncogenesis plausibility. Consequently, the expert evaluations cannot be viewed as the product of a single uniform institutional framework for critical appraisal. While it is conceivable that some shared perspectives could arise through a common institutional environment, given the vast and heterogeneous data sources used to train contemporary LLMs, it is unlikely that agreement with the expert panel resulted from exposure to appraisal patterns specific to a single South African institution. It is therefore more plausible that the observed alignment between GPT-5 models and experts reflects the models' ability to identify broadly recognized indicators of paper reliability and strength of evidence than the replication of highly specific institutional appraisal norms.

While GPT-5 models appeared to display evidence appraisal capabilities comparable to experts, calibration or multi-model strategies may still be necessary when deploying LLMs in evaluative biomedical workflows requiring evidence appraisal. GPT-5 models remained more stringent in aspects of critical evaluation (tended to provide slightly lower ratings for paper reliability score and strength of evidence score than experts), while simultaneously failing to identify issues detected by human evaluators, including contradictions and methodological concerns. Thus, similarity in overall appraisal scores did not necessarily correspond to human-level identification of specific weaknesses within a study. Future studies could combine models with complementary strengths, for example use a model specifically fine-tuned for research quality assessment or trustworthiness evaluation to identify potential limitations, sources of bias, and methodological concerns, and then incorporate these findings into the appraisal process of a more generalist model such as GPT-5. Such architectures could potentially leverage the human-like scoring behavior observed in frontier LLMs while improving the depth and consistency of evidence appraisal, resulting in evaluations that more closely approximate expert review.

\subsection{Applying Microbial Oncogenesis Criteria}
While Gemini 2.5 Pro, Gemini 2.5 Flash, and GPT-5 all demonstrated superior ability in extracting evidence to support and/or refute microbial oncogenesis criteria (adapted from \cite{vanDorsten2025}), GPT-5 applied and scored these criteria most similarly to experts, thereby producing cumulative oncogenesis scores comparable to experts (Fig.~\ref{fig:cumulative_oncogenesis_mcp_impact}A). GPT-5 had one instance of disagreement with expert consensus, but its reasoning was sound and indicated strict adherence to the provided microbial causal criteria definitions, increasing trust in its reasoning and consistency for future outputs. GPT-5 Nano appeared to apply and score the microbial oncogenesis criteria similarly to experts, yet had instances of inconsistent reasoning (see Sec.~\ref{sec:omissions}), in addition to less thorough evidence extraction than the other models (described in Sec.~\ref{sec:omissions} and Sec.~\ref{sec:providing_trustworthy_outputs}). Gemini 2.5 Pro and Gemini 2.5 Flash applied the criteria leniently, stating that criteria were fulfilled more often than experts did, thereby significantly decreasing inter-expert agreement on cumulative oncogenesis scores (Fig.~\ref{fig:cumulative_oncogenesis_mcp_impact}A). Further analysis found this leniency to be highly inappropriate, with susceptibility to ecological inference bias and false-cause reasoning, in addition to claims that indirect associative evidence (e.g., reviews of publicly accessible prevalence data) fulfilled multiple criteria requiring experimental or clinical evidence (e.g., temporality, dose-response relationship, and prevention).

Overall, Gemini 2.5 Pro, Gemini 2.5 Flash, and GPT-5 all demonstrated superior ability to extract evidence for nuanced microbial oncogenesis criteria to the level of domain experts, while GPT-5 alone applied the criteria most consistently and to expert standards. LLMs are therefore capable of applying and extracting evidence for nuanced criteria such as these, yet levels of competency and consistency in these tasks differ between models, requiring piloting and calibration of these tools before deployment. However, smaller generalist LLMs, such as GPT-5 Nano, may lack the domain knowledge and reasoning complexity to consistently apply and extract sufficient evidence for these criteria.

\subsection{Producing Accurate, Trustworthy Outputs with Minimal Hallucinations or Omissions}
\label{sec:providing_trustworthy_outputs}
Contrary to widespread concerns regarding prevalence of hallucinations in biomedical applications \cite{Kim2025,Artsi2025}, these were infrequent and largely restricted to smaller models, with most instances representing misinterpretations of the text (distortions) rather than typical fabrications (hallucinations). The majority of instances in which LLMs introduced information absent from expert responses represented paper-supported additions derived from the source paper rather than hallucinations or distortions (Table~\ref{tab:hallucinations_omissions_accurate}). LLMs more frequently had omissions in their responses, with most reflecting differences in emphasis or granularity rather than substantive misunderstanding. However, in two oncogenesis criteria question instances, GPT-5 and GPT-5 Nano failed to classify criteria as fulfilled where experts unanimously did so, suggesting potential rigidity in microbial oncogenicity criteria interpretation by these models (see Sec.~\ref{sec:omissions}). Conversely, LLMs frequently included paper-supported additions not identified by individual experts, suggesting differences in answer focus rather than inferential error. These low hallucination rates suggest that structured extraction templates, explicit criteria definitions, and constrained task framing may substantially mitigate hallucination risk. This is consistent with emerging evidence that structured prompting and reasoning scaffolds improve factual reliability \cite{cheng-etal-2025-chain, Anh-Hoang2025}, while other prompt refinements, including increased question specificity and minimum detail requirements, may reduce the frequency of omissions in future prompt iterations \cite{Singhal2023,Kim2025}. 

Notably, hallucinations and distortions were isolated to two question groups: (1) methodology-focused questions, which required identification of study limitations and statistical interpretation, and (2) contradiction identification, which required understanding of nuanced biomedical language and interpretation of tables and figures. 

Methodological critique demands integration of contextual domain knowledge (which generalist LLMs - especially smaller models - may lack), statistical reasoning, and inferential judgment, all of which are areas in which LLM limitations have been previously observed \cite{Kim2025,Boye2025}. Presently, when asked to identify potential limitations of studies, all LLMs tended to repeat the same set of limitations with minor changes to ensure contextual appropriateness, and were less likely than experts to comment on potential errors or information gaps. These failures illustrate the limits of domain-specific knowledge and skills in these generalist models, yet, there is clear potential to boost model performance in this regard, with instances where LLMs noted pivotal limitations that experts missed, and models tending to capture limitations from more than one expert in their responses. Overall, while integration of biomedical domain-specific models in a multi-model system may improve methodological appraisal ability, generalist model fine-tuning and prompt modifications may offer a similar performance boost while retaining the advanced reasoning abilities seen in these generalist models.

Poor performance in identifying contradictions within the papers was unsurprising, considering prior evidence that LLMs have difficulty identifying contradictions in provided full-text papers \cite{Li2024} and isolated sentence pairs or text segments, with similar performance across models despite size differences in zero-shot settings \cite{Tam2023}. This difficulty is not unique to LLMs, with studies indicating that humans struggle to identify in-text contradictions, from peer-reviewers failing to identify inconsistencies between paper abstracts and full report results \cite{Li2017}, and studies demonstrating human difficulty in identifying contradictions in both full-text documents \cite{Li2024} and isolated paragraphs, even when provided text segments are short \cite{Otero1992}. In contrast, when given the isolated paper sections containing the human-identified contradictions, LLMs performed considerably better, identifying close to all human-identified contradictions and detecting additional contradictions within these sections that the human evaluator missed. This indicates that while LLMs may have the ability to identify contradictions in research papers, current context window limitations likely preclude usability in this regard for full-text papers. This ``context rot" poses an issue for our proposed pipeline, which relies on full-text papers. This requires the development of novel methods to circumvent large context size limitations and enable identification of errors for thorough evaluation of paper reliability and strength of evidence.

Overall, frontier LLMs' outputs appear to be sufficiently accurate to be trusted in evidence synthesis workflows, demonstrating comprehensiveness with minimal hallucinations. Although omissions were frequent, their impact was predominantly minor, with LLMs still tending to include information spanning across the three experts responses rather than aligning with singular experts. Smaller LLMs' outputs were less trustworthy, with higher (although still limited) instances of hallucinations and/or distortions, likely due to their poorer ability to interpret nuanced biomedical language compared to frontier LLMs. Methodology-related questions and contradiction detection remain the greatest areas of vulnerability for generalist LLMs due to domain knowledge and context window limitations.

\subsection{Limitations}
This study has several limitations. First, evaluation was restricted to a single MCP. This MCP was deliberately selected because of its unconfirmed status, conflicting evidence base, and predominance of associative rather than causal evidence, characteristics that are likely to resemble the intended real-world application of the system and many future MCPs of interest. Nevertheless, a single MCP cannot capture the full spectrum of evidence profiles encountered across potential MCPs. Although there is no obvious reason to expect the evaluated LLMs to be systematically biased towards or against this specific MCP, additional validation across MCPs with differing levels of evidential support would potentially strengthen confidence in the generalizability of these findings. In particular, future studies could evaluate MCPs approaching confirmed oncogenic status, including papers that challenge their plausibility, to assess whether model prior knowledge influences evaluation. Similarly, testing MCPs supported by sparse or emerging evidence could help determine whether model performance extends to more novel hypotheses. Consistent performance across such diverse MCPs could provide stronger evidence for the robustness and broader applicability of the proposed pipeline. Additionally, while the focus of this study is on applying this system to evaluating evidence on MCPs specifically, the choice of an MCP may limit generalizability of findings to other domains. 

Second, the dataset comprised 24 papers, reflecting the intensive nature of the expert annotation process. While the extraction template consisted of 77 questions, allowing for extensive points of analysis across this dataset, 24 papers remain a small sample. Completion of the extraction template typically required more than two hours per paper, making large-scale expert evaluation impractical. Although this constrained the dataset size, it mirrors the real-world challenge that motivated the present work: the volume of biomedical literature far exceeds the capacity for comprehensive manual review. The ability of LLMs to perform comparable evaluations within minutes suggests a potential route to scaling such analyses, with experts serving in oversight and validation roles rather than conducting every assessment manually.

Third, the cohort of seven experts that provided responses for the human-validated test dataset were recruited from a single academic institution, and only three experts were assigned to each paper. While the study design of having independent review and extraction of information from papers was intentional to mimic current systematic evidence synthesis methods, a Delphi panel for consensus may have refined expert responses and further improved inter-expert agreement. Fourth, while the creation and use of a structured extraction template was intentional to standardize comparisons and enable control and transparency of LLM outputs, its use inevitably affects LLM reasoning patterns, which may improve model performance and reduce generalizability of findings to studies or domains that do not use such structured prompting. 

Finally, LLM developers make frequent changes to their models, including updates to existing, already released models (e.g., updates to the Gemini 2.5 Pro model without access to previous versions), and further releases of brand new models (e.g., the release of GPT-5.1, a distinct model from GPT-5). Updates to existing models can result in these LLMs having different abilities and competencies than when they were originally tested, including potential decreases in performance. The models included in this study are considered stable due to shifted focus to and release of newer model ranges by Google and OpenAI (e.g., Gemini 3 and GPT-5.5 model ranges), which reduces concerns regarding changes in the included models' performance in future. However, as newer models are released, the potential for older ones to eventually be depreciated (i.e., no longer accessible) remains. Furthermore, the included older models may fail to represent the full spectrum of these newer frontier LLMs' abilities in the various discussed domains. However, our aim was to determine whether current LLMs were generally competent and trustworthy enough to be used for automated systematic evidence synthesis purposes, which we have found to be true. Each newly released LLM tends to beat prior models' performance, and if this trend continues, the concerns found in the present study may no longer be applicable to these new models.

\subsection{Conclusion}
Overall, pre-trained reasoning LLMs (particularly GPT-5 models) were indistinguishable from experts across the structured biomedical evidence extraction and appraisal tasks. Hallucinations were rare in larger models (Gemini 2.5 Pro, GPT-5) under these constrained prompting conditions, but more frequent in smaller models (Gemini 2.5 Flash, GPT-5 Nano), although the rate of these errors could not be directly compared with human experts. These findings support integration of reasoning LLMs into AI evidence synthesis systems, particularly for evidence extraction and integrative summarization from individual biomedical research papers. However, methodological critique, scoring biases, and identification of contradictions remain key areas requiring further system refinement.

\section*{Conflict of Interest Statement}
The authors declare that the research was conducted in the absence of any commercial or financial relationships that could be construed as a potential conflict of interest.

\section*{Author Contributions}
KK: Conceptualization, Methodology, Investigation, Data curation, Formal analysis, Writing - original draft.
BAB, RFB, RK, MZM, EKS, and HW: Conceptualization, Methodology, Investigation, Software (development of the data collection application, by EKS), Supervision (BAB, RFB, and RK), and contribution to the development and refinement of the evaluation framework, including hallucination classification; Writing - review \& editing.
RD, NI, NAI, KN, JN, EEN, and RP: Investigation and Data curation (expert dataset generation); Writing - review \& editing.
All authors contributed to manuscript revision, read, and approved the submitted version.

\section*{Funding}
The work reported herein was made possible through funding by the South African Medical Research Council (SAMRC) through its Division of Research Capacity Development under the SAMRC Clinician Researcher Development Programme with funding received from the National Department of Health.
Additional support was provided by the Wits Health Consortium and the Infectious Diseases and Oncology Research Institute (IDORI). Article processing charges for this publication were supported by IDORI.

\section*{Acknowledgments}
The authors thank Prof. Raquel Duarte and Ms Caryn McNamara for their advice and support on this project.

\newpage
\bibliographystyle{plain}
\bibliography{ref}

\begin{thebibliography}{10}

\bibitem{NatureReviewsMicrobiology2011}
Microbiology by numbers.
\newblock {\em Nature Reviews Microbiology}, 9:628--628, 8 2011.

\bibitem{Anh-Hoang2025}
Dang Anh-Hoang, Vu~Tran, and Le~Minh Nguyen.
\newblock Survey and analysis of hallucinations in large language models: attribution to prompting strategies or model behavior.
\newblock {\em Frontiers in Artificial Intelligence}, 8:1622292, 9 2025.

\bibitem{Artsi2025}
Yaara Artsi, Vera Sorin, Benjamin~S. Glicksberg, Panagiotis Korfiatis, Robert Freeman, Girish~N. Nadkarni, and et~al.
\newblock Challenges of implementing llms in clinical practice: Perspectives.
\newblock {\em Journal of Clinical Medicine}, 14:6169, 9 2025.

\bibitem{Bernardo2023}
Giancarla Bernardo, Valentino~Le Noci, Martina~Di Modica, Elena Montanari, Tiziana Triulzi, Serenella~M. Pupa, and et~al.
\newblock The emerging role of the microbiota in breast cancer progression.
\newblock {\em Cells}, 12:1945, 8 2023.

\bibitem{Boye2025}
Johan Boye and Birger Moell.
\newblock Large language models and mathematical reasoning failures.
\newblock 2 2025.

\bibitem{Brigo2025}
Francesco Brigo, Serena Broggi, Gionata Strigaro, Sasha Olivo, Valentina Tommasini, Magdalena Massar, and et~al.
\newblock Artificial intelligence (chatgpt 4.0) vs. human expertise for epileptic seizure and epilepsy diagnosis and classification in adults: An exploratory study.
\newblock {\em Epilepsy and Behavior}, 166, 5 2025.

\bibitem{Blbl2025}
Ogün Bülbül, Hande~Melike Bülbül, and Esat Kaba.
\newblock Assessing chatgpt’s summarization of 68ga psma pet/ct reports for patients.
\newblock {\em Abdominal Radiology}, 50:1467--1474, 3 2025.

\bibitem{Callahan2012}
Robert Callahan, Uma Mudunuri, Sharon Bargo, Ahmed Raafat, David Mccurdy, Corinne Boulanger, and et~al.
\newblock Genes affected by mouse mammary tumor virus (mmtv) proviral insertions in mouse mammary tumors are deregulated or mutated in primary human mammary tumors.
\newblock {\em Oncotarget}, 3:1320, 2012.

\bibitem{Castelvecchi2016}
Davide Castelvecchi.
\newblock Can we open the black box of ai?
\newblock {\em Nature News}, 538:20, 10 2016.

\bibitem{Cedro-Tanda2014}
Alberto Cedro-Tanda, Alejandro Córdova-Solis, Teresa Juárez-Cedillo, Emmanuel Pina-Jiménez, Marta~E. Hernández-Caballero, Christian Moctezuma-Meza, and et~al.
\newblock Prevalence of hmtv in breast carcinomas and unaffected tissue from mexican women.
\newblock {\em BMC cancer}, 14:942, 12 2014.

\bibitem{Chappell2023}
Mary Chappell, Mary Edwards, Deborah Watkins, Christopher Marshall, and Sara Graziadio.
\newblock Machine learning for accelerating screening in evidence reviews.
\newblock {\em Cochrane Evidence Synthesis and Methods}, 1:e12021, 7 2023.

\bibitem{cheng-etal-2025-chain}
Jiahao Cheng, Tiancheng Su, Jia Yuan, Guoxiu He, Jiawei Liu, Xinqi Tao, Jingwen Xie, and Huaxia Li.
\newblock Chain-of-thought prompting obscures hallucination cues in large language models: An empirical evaluation.
\newblock In Christos Christodoulopoulos, Tanmoy Chakraborty, Carolyn Rose, and Violet Peng, editors, {\em Findings of the Association for Computational Linguistics: EMNLP 2025}, pages 1272--1305, Suzhou, China, November 2025. Association for Computational Linguistics.

\bibitem{Comanici2025}
Gheorghe Comanici, Eric Bieber, Mike Schaekermann, Ice Pasupat, Noveen Sachdeva, Inderjit Dhillon, and et~al.
\newblock Gemini 2.5: Pushing the frontier with advanced reasoning, multimodality, long context, and next generation agentic capabilities.
\newblock 7 2025.

\bibitem{Croxford2025}
Emma Croxford, Yanjun Gao, Elliot First, Nicholas Pellegrino, Miranda Schnier, John Caskey, Madeline Oguss, Graham Wills, Guanhua Chen, Dmitriy Dligach, Matthew~M. Churpek, Anoop Mayampurath, Frank Liao, Cherodeep Goswami, Karen~K. Wong, Brian~W. Patterson, and Majid Afshar.
\newblock Evaluating clinical ai summaries with large language models as judges.
\newblock {\em npj Digital Medicine 2025 8:1}, 8:640--, 11 2025.

\bibitem{Dawood2025}
Muhammed~Muaaz Dawood, Mohammad~Zaid Moonsamy, Kaela Kokkas, Hairong Wang, Robert~F. Breiman, Richard Klein, and et~al.
\newblock Small language models can use nuanced reasoning for health science research classification: A microbial-oncogenesis case study.
\newblock 12 2025.

\bibitem{deMartel2020}
Catherine de~Martel, Damien Georges, Freddie Bray, Jacques Ferlay, and Gary~M. Clifford.
\newblock Global burden of cancer attributable to infections in 2018: a worldwide incidence analysis.
\newblock {\em The Lancet Global Health}, 8:e180--e190, 2 2020.

\bibitem{Pereira2020}
Nathália de~Sousa~Pereira, Glauco Akelinghton~Freire Vitiello, Bruna~Karina Banin-Hirata, Glaura Scantamburlo~Alves Fernandes, Maria José~Sparça Salles, and et~al.
\newblock Mouse mammary tumor virus (mmtv)-like env sequence in brazilian breast cancer samples: Implications in clinicopathological parameters in molecular subtypes.
\newblock {\em International journal of environmental research and public health}, 17:1--14, 12 2020.

\bibitem{AlDossary2018}
Reem~Al Dossary, Khaled~R. Alkharsah, and Haitham Kussaibi.
\newblock Prevalence of mouse mammary tumor virus (mmtv)-like sequences in human breast cancer tissues and adjacent normal breast tissues in saudi arabia.
\newblock {\em BMC cancer}, 18:170, 2 2018.

\bibitem{Dougherty2023}
Michael~W. Dougherty and Christian Jobin.
\newblock Intestinal bacteria and colorectal cancer: etiology and treatment.
\newblock {\em Gut Microbes}, 15:2185028, 3 2023.

\bibitem{Ghasemi2022}
Asghar Ghasemi, Parvin Mirmiran, Khosrow Kashfi, and Zahra Bahadoran.
\newblock Scientific publishing in biomedicine: A brief history of scientific journals.
\newblock {\em International Journal of Endocrinology and Metabolism}, 21:e131812, 1 2022.

\bibitem{Goedert2006}
J.~J. Goedert, C.~S. Rabkin, and S.~R. Ross.
\newblock Prevalence of serologic reactivity against four strains of mouse mammary tumour virus among us women with breast cancer.
\newblock {\em British Journal of Cancer}, 94:548--551, 2 2006.

\bibitem{Gottweis2026}
Juraj Gottweis, Wei-Hung Weng, Alexander Daryin, Tao Tu, Petar Sirkovic, Artiom Myaskovsky, Grzegorz Glowaty, Felix Weissenberger, Alessio Orlandi, Dan Popovici, Anil Palepu, Keran Rong, Ryutaro Tanno, Khaled Saab, Fan Zhang, Jacob Blum, Andrew Carroll, Kavita Kulkarni, Nenad Tomašev, Dina Zverinski, Ivor Rendulic, Elahe Vedadi, Florian Hasler, Luka Rimanic, Marina Boia, Ivan Budiselic, Ben Feinstein, Mathias Bellaiche, Tom Sheffer, Jan Freyberg, Jeremy Ratcliff, Ottavia Bertolli, Katherine Chou, Avinatan Hassidim, Burak Gokturk, Amin Vahdat, Yuan Guan, Vikram Dhillon, Eeshit~Dhaval Vaishnav, Byron Lee, Tiago R.~D. Costa, José~R. Penadés, Gary Peltz, Yossi Matias, James Manyika, Demis Hassabis, Yunhan Xu, Pushmeet Kohli, Annalisa Pawlosky, Alan Karthikesalingam, and Vivek Natarajan.
\newblock Accelerating scientific discovery with co-scientist.
\newblock {\em Nature 2026 655:8122}, 655:487--496, 5 2026.

\bibitem{Gough2020}
David Gough, Phil Davies, Gro Jamtvedt, Etienne Langlois, Julia Littell, Tamara Lotfi, and et~al.
\newblock Evidence synthesis international (esi): Position statement.
\newblock {\em Systematic Reviews 2020 9:1}, 9:155--, 7 2020.

\bibitem{Gupta2022}
Ishita Gupta, Reem Al-Sarraf, Hanan Farghaly, Semir Vranic, Ali~A. Sultan, Hamda Al-Thawadi, and et~al.
\newblock Incidence of hpvs, ebv, and mmtv-like virus in breast cancer in qatar.
\newblock {\em Intervirology}, 65:188--194, 10 2022.

\bibitem{Gupta2021}
Ishita Gupta, Monika Ulamec, Melita Peric-Balja, Snjezana Ramic, Ala Eddin~Al Moustafa, Semir Vranic, and et~al.
\newblock Presence of high-risk hpvs, ebv, and mmtv in human triple-negative breast cancer.
\newblock {\em Human vaccines \& immunotherapeutics}, 17:4457--4466, 2021.

\bibitem{Huang2025}
Lei Huang, Weijiang Yu, Weitao Ma, Weihong Zhong, Zhangyin Feng, Haotian Wang, and et~al.
\newblock A survey on hallucination in large language models: Principles, taxonomy, challenges, and open questions.
\newblock {\em ACM Trans. Inf. Syst.}, 43(2):42, January 2025.

\bibitem{IARC2012}
{IARC Working Group on the Evaluation of Carcinogenic Risks to Humans}.
\newblock {\em General Remarks}, volume 100B, page~35.
\newblock {International Agency for Research on Cancer (IARC)}, 2012.

\bibitem{Indik2007}
Stanislav Indik, Walter~H. Günzburg, Pavel Kulich, Brian Salmons, and Francoise Rouault.
\newblock Rapid spread of mouse mammary tumor virus in cultured human breast cells.
\newblock {\em Retrovirology}, 4:73, 10 2007.

\bibitem{InternationalAgencyforResearchonCancerIARC2025}
{International Agency for Research on Cancer (IARC)}.
\newblock List of classifications: Agents classified by the iarc monographs, volumes 1-137, 2025.

\bibitem{James2024}
Lisa~M. James and Apostolos~P. Georgopoulos.
\newblock Breast cancer, viruses, and human leukocyte antigen (hla).
\newblock {\em Scientific reports}, 14:16179, 12 2024.

\bibitem{Jumper2021}
John Jumper, Richard Evans, Alexander Pritzel, Tim Green, Michael Figurnov, Olaf Ronneberger, and et~al.
\newblock Highly accurate protein structure prediction with alphafold.
\newblock {\em Nature 2021 596:7873}, 596:583--589, 7 2021.

\bibitem{FawadKhalid2021}
Hafiz~Fawad Khalid, Amjad Ali, Nida Fawad, Shazia Rafique, Inam Ullah, Gouhar Rehman, and et~al.
\newblock Mmtv-like virus and c-myc over-expression are associated with invasive breast cancer.
\newblock {\em Genetics and Evolution}, 91:104827, 2021.

\bibitem{Khalid2023}
Hafiz~Fawad Khalid, Sadia Bibi, Amjad Ali, Nida Fawad, Muhammad~Usman Shams, Wafa Idrees, and et~al.
\newblock Decoding the mystery of mmtv-like virus and its relationship with breast cancer metastasis.
\newblock {\em Journal of infection and public health}, 16:1396--1402, 9 2023.

\bibitem{Kim2025}
Yubin Kim, Hyewon Jeong, Shan Chen, Shuyue~Stella Li, Chanwoo Park, Mingyu Lu, and et~al.
\newblock Medical hallucinations in foundation models and their impact on healthcare.
\newblock 11 2025.

\bibitem{Kincaid2018}
Rodney~P. Kincaid, Neena~G. Panicker, Mary~M. Lozano, Christopher~S. Sullivan, Jaquelin~P. Dudley, and Farah Mustafa.
\newblock Mmtv does not encode viral micrornas but alters the levels of cancer-associated host micrornas.
\newblock {\em Virology}, 513:180--187, 1 2018.

\bibitem{Koh2025}
Matthew Chung~Yi Koh, Jinghao~Nicholas Ngiam, Jolene Ee~Ling Oon, Lionel Hon~Wai Lum, Nares Smitasin, and Sophia Archuleta.
\newblock Using chatgpt for writing hospital inpatient discharge summaries – perspectives from an inpatient infectious diseases service.
\newblock {\em BMC Health Services Research}, 25:221, 12 2025.

\bibitem{Kulkarni2013}
Bhushan~B. Kulkarni, Shivaprakash~V. Hiremath, Suyamindra~S. Kulkarni, Umesh~R. Hallikeri, Basavaraj~R. Patil, and Pramod~B. Gai.
\newblock Genomic dna of mcf-7 breast cancer cells not an ideal choice as positive control for pcr amplification based detection of mouse mammary tumor virus-like sequences.
\newblock {\em Journal of Virological Methods}, 193:304--307, 11 2013.

\bibitem{Lessi2020}
Francesca Lessi, Nicole Grandi, Chiara~Maria Mazzanti, Prospero Civita, Cristian Scatena, Paolo Aretini, and et~al.
\newblock A human mmtv-like betaretrovirus linked to breast cancer has been present in humans at least since the copper age.
\newblock {\em Aging}, 12:15978--15994, 8 2020.

\bibitem{Li2017}
Guowei Li, Luciana~P.F. Abbade, Ikunna Nwosu, Yanling Jin, Alvin Leenus, Muhammad Maaz, and et~al.
\newblock A scoping review of comparisons between abstracts and full reports in primary biomedical research.
\newblock {\em BMC medical research methodology}, 17:181, 12 2017.

\bibitem{Li2024}
Jierui Li, Vipul Raheja, and Dhruv Kumar.
\newblock Contradoc: Understanding self-contradictions in documents with large language models.
\newblock {\em Proceedings of the 2024 Conference of the North American Chapter of the Association for Computational Linguistics: Human Language Technologies, NAACL 2024}, 1:6509--6523, 2024.

\bibitem{Li2025}
Yilan Li, Tianshu Gu, Chengyuan Yang, Minghui Li, Congyi Wang, Lan Yao, and et~al.
\newblock Ai-assisted hypothesis generation to address challenges in cardiotoxicity research: Simulation study using chatgpt with gpt-4o.
\newblock {\em Journal of Medical Internet Research}, 27:e66161, 2025.

\bibitem{Luczynski2022}
Pauline Luczynski, Philip Poulin, Kamila Romanowski, and James~C. Johnston.
\newblock Tuberculosis and risk of cancer: A systematic review and meta-analysis.
\newblock {\em PLOS ONE}, 17:e0278661, 12 2022.

\bibitem{Marzi2024}
Giacomo Marzi, Marco Balzano, and Davide Marchiori.
\newblock K-alpha calculator–krippendorff's alpha calculator: A user-friendly tool for computing krippendorff's alpha inter-rater reliability coefficient.
\newblock {\em MethodsX}, 12:102545, 6 2024.

\bibitem{Mazzanti2015}
Chiara~Maria Mazzanti, Francesca Lessi, Ivana Armogida, Katia Zavaglia, Sara Franceschi, Mohammad~Al Hamad, and et~al.
\newblock Human saliva as route of inter-human infection for mouse mammary tumor virus.
\newblock {\em Oncotarget}, 6:18355--18363, 2015.

\bibitem{McFadden2023}
Benjamin~R. McFadden, Mark Reynolds, and Timothy~J.J. Inglis.
\newblock Developing machine learning systems worthy of trust for infection science: a requirement for future implementation into clinical practice.
\newblock {\em Frontiers in Digital Health}, 5:1260602, 9 2023.

\bibitem{Miyabayashi2022}
Koji Miyabayashi, Hideaki Ijichi, and Mitsuhiro Fujishiro.
\newblock The role of the microbiome in pancreatic cancer.
\newblock {\em Cancers}, 14:4479, 9 2022.

\bibitem{Moore2010}
Patrick~S. Moore and Yuan Chang.
\newblock Why do viruses cause cancer? highlights of the first century of human tumour virology.
\newblock {\em Nature reviews. Cancer}, 10:878, 12 2010.

\bibitem{Morales-Snchez2013}
Abigail Morales-Sánchez, Tzindilú Molina-Muñoz, Juan~L.E. Martínez-López, Paulina Hernández-Sancén, Alejandra Mantilla, Yelda~A. Leal, and et~al.
\newblock No association between epstein-barr virus and mouse mammary tumor virus with breast cancer in mexican women.
\newblock {\em Scientific reports}, 3, 10 2013.

\bibitem{Naccarato2019}
Antonio~Giuseppe Naccarato, Francesca Lessi, Katia Zavaglia, Cristian Scatena, Mohammad A.~Al Hamad, Paolo Aretini, and et~al.
\newblock Mouse mammary tumor virus (mmtv) - like exogenous sequences are associated with sporadic but not hereditary human breast carcinoma.
\newblock {\em Aging}, 11:7236--7241, 9 2019.

\bibitem{NationalCancerInstituteAZ}
{National Cancer Institute}.
\newblock A to z list of cancer types.

\bibitem{NationalCancerInstitute2021}
{National Cancer Institute}.
\newblock What is cancer?, 10 2021.
\newblock Accessed: 13/08/2025.

\bibitem{Naushad2017}
Wasifa Naushad, Orooj Surriya, and Hajra Sadia.
\newblock Prevalence of ebv, hpv and mmtv in pakistani breast cancer patients: A possible etiological role of viruses in breast cancer.
\newblock {\em Infection, genetics and evolution : journal of molecular epidemiology and evolutionary genetics in infectious diseases}, 54:230--237, 10 2017.

\bibitem{Nordmann2025}
Kim Nordmann, Stefanie Sauter, Mirjam Stein, Johanna Aigner, Marie-Christin Redlich, Michael Schaller, and et~al.
\newblock Evaluating the performance of artificial intelligence in summarizing pre-coded text to support evidence synthesis: a comparison between chatbots and humans.
\newblock {\em BMC Medical Research Methodology}, 25:150, 2025.

\bibitem{Otero1992}
José Otero and Walter Kintsch.
\newblock Failures to detect contradictions in a text: What readers believe versus what they read.
\newblock {\em Psychological Science (0956-7976)}, 3:229--235, 7 1992.

\bibitem{Otten1988}
Anne~D. Otten, Michel~M. Sanders, and G.~Stanley McKnight.
\newblock The mmtv ltr promoter is induced by progesterone and dihydrotestosterone but not by estrogen.
\newblock {\em Molecular endocrinology (Baltimore, Md.)}, 2:143--147, 1988.

\bibitem{Parkin2020}
Donald~M. Parkin, Lucia Hämmerl, Jacques Ferlay, and Eva~J. Kantelhardt.
\newblock Cancer in africa 2018: The role of infections.
\newblock {\em International Journal of Cancer}, 146:2089--2103, 4 2020.

\bibitem{Perzova2017}
Raisa Perzova, Lynn Abbott, Patricia Benz, Steve Landas, Seema Khan, Jordan Glaser, and et~al.
\newblock Is mmtv associated with human breast cancer? maybe, but probably not.
\newblock {\em Virology journal}, 14:196, 10 2017.

\bibitem{Reese2026}
May~Lynn Reese, Markela Zeneli, Mindy Ng, Jacob Haimes, Andreea Damien, and Elizabeth Stade.
\newblock Using llm-as-a-judge/jury to advance scalable, clinically-validated safety evaluations of model responses to users demonstrating psychosis.
\newblock {\em Proceedings of IASEAI Conference}, 2:610--624, 7 2026.

\bibitem{Reza2015}
Malekpour~Afshar Reza, Mollaie~Hamid Reza, Lashkarizadeh Mahdiyeh, Fazlalipour Mehdi, and Zeinali~Nejad Hamid.
\newblock Evaluation frequency of merkel cell polyoma, epstein-barr and mouse mammary tumor viruses in patients with breast cancer in kerman, southeast of iran.
\newblock {\em Asian Pacific journal of cancer prevention : APJCP}, 16:7351--7357, 2015.

\bibitem{Shariatpanahi2017}
Shervin Shariatpanahi, Najma Farahani, Ahmad~Reza Salehi, and Rasoul Salehi.
\newblock High prevalence of mouse mammary tumor virus-like gene sequences in breast cancer samples of iranian women.
\newblock {\em Nucleosides, nucleotides \& nucleic acids}, 36:621--630, 2017.

\bibitem{Siam2025}
Md~Kamrul Siam, Angel Varela, Md~Jobair~Hossain Faruk, Jerry~Q. Cheng, Huanying Gu, Abdullah~Al Maruf, and et~al.
\newblock Benchmarking large language models on the united states medical licensing examination for clinical reasoning and medical licensing scenarios.
\newblock {\em Scientific Reports 2025 16:1}, 16:1387--, 12 2025.

\bibitem{singh2026openaigpt5card}
Aaditya Singh, Adam Fry, Adam Perelman, Adam Tart, Adi Ganesh, Ahmed El-Kishky, and et~al.
\newblock Openai gpt-5 system card, 2026.

\bibitem{Singhal2023}
Karan Singhal, Shekoofeh Azizi, Tao Tu, S.~Sara Mahdavi, Jason Wei, Hyung~Won Chung, and et~al.
\newblock Large language models encode clinical knowledge.
\newblock {\em Nature 2023 620:7972}, 620:172--180, 7 2023.

\bibitem{Stewart2022}
Alexandre~F.R. Stewart and Hsiao~Huei Chen.
\newblock Revisiting the mmtv zoonotic hypothesis to account for geographic variation in breast cancer incidence.
\newblock {\em Viruses}, 14:559, 3 2022.

\bibitem{Tabriz2013}
Hedieh~Moradi Tabriz, Kazem Zendehdel, Reza Shahsiah, Forouzandeh Fereidooni, Baharak Mehdipour, and Zahra~Mostakhdemin Hosseini.
\newblock Lack of detection of the mouse mammary tumor-like virus (mmtv) env gene in iranian women breast cancer using real time pcr.
\newblock {\em Asian Pacific journal of cancer prevention : APJCP}, 14:2945--2948, 2013.

\bibitem{Tam2023}
Derek Tam, Anisha Mascarenhas, Shiyue Zhang, Sarah Kwan, Mohit Bansal, and Colin Raffel.
\newblock Evaluating the factual consistency of large language models through news summarization.
\newblock {\em Proceedings of the Annual Meeting of the Association for Computational Linguistics}, Findings of the Association for Computational Linguistics: ACL 2023:5220--5255, 2023.

\bibitem{Tsafnat2014}
Guy Tsafnat, Paul Glasziou, Miew~K. Choong, Adam Dunn, Filippo Galgani, and Enrico Coiera.
\newblock Systematic review automation technologies.
\newblock {\em Systematic Reviews}, 3:1--15, 4 2014.

\bibitem{vandeSchoot2021}
Rens van~de Schoot, Jonathan de~Bruin, Raoul Schram, Parisa Zahedi, Jan de~Boer, Felix Weijdema, and et~al.
\newblock An open source machine learning framework for efficient and transparent systematic reviews.
\newblock {\em Nature Machine Intelligence 2021 3:2}, 3:125--133, 2 2021.

\bibitem{vanDorsten2025}
Rebecca~Toumi van Dorsten and Robert~F Breiman.
\newblock A landscape review with novel criteria to evaluate microbial drivers for cancer: Priorities for innovative research targeting excessive cancer mortality in sub-saharan africa.
\newblock {\em Frontiers in Cellular and Infection Microbiology}, 15, 7 2025.

\bibitem{Watts2023}
Geoff Watts.
\newblock Harald zur hausen.
\newblock {\em The Lancet}, 402:20, 7 2023.

\bibitem{Wei2022}
Jason Wei, Xuezhi Wang, Dale Schuurmans, Maarten Bosma, Brian Ichter, Fei Xia, and et~al.
\newblock Chain-of-thought prompting elicits reasoning in large language models.
\newblock {\em Advances in Neural Information Processing Systems}, 35:24824--24837, 1 2022.

\bibitem{WorldHealthOrganisationWHO2022}
{World Health Organisation (WHO)}.
\newblock Cervical cancer elimination initiative, 2022.
\newblock Accessed: 14/01/2025.

\bibitem{Zambaldi2024}
Vinicius Zambaldi, David La, Alexander~E Chu, Harshnira Patani, Amy~E Danson, Tristan O~C Kwan, and et~al.
\newblock De novo design of high-affinity protein binders with alphaproteo.
\newblock 9 2024.

\bibitem{BouZerdan2022}
Maroun~Bou Zerdan, Joseph Kassab, Paul Meouchy, Elio Haroun, Rami Nehme, Morgan~Bou Zerdan, and et~al.
\newblock The lung microbiota and lung cancer: A growing relationship.
\newblock {\em Cancers}, 14:4813, 10 2022.

\bibitem{Zhang2025}
Kang Zhang, Xin Yang, Yifei Wang, Yunfang Yu, Niu Huang, Gen Li, and et~al.
\newblock Artificial intelligence in drug development.
\newblock {\em Nature Medicine}, 31:45--59, 1 2025.

\end{thebibliography}

\newpage
\appendix
\section{Supplementary Material}
\setcounter{figure}{0}
\setcounter{table}{0}
\renewcommand{\thefigure}{S\arabic{figure}}
\renewcommand{\thetable}{S\arabic{table}}

\subsection{Figures}

\begin{figure}[h]
    \caption{Assessment prompt for LLM judge (GPT-5, high reasoning effort) for pairwise long-answer comparisons (pairwise overlap scores).}
    \label{fig:1v1_assessment_prompt}
    \begin{framed}
        You are serving as an impartial medical research evaluator. Your task is to score how well two responses match each other in meaning and nuance. The topic is whether HMTV / MMTV-like virus can cause breast cancer in humans.\\
\\
\textbf{SCORING RUBRIC:}
\begin{itemize}
    \item \quad 0: No overlap: Responses share no common points.
    \item \quad  1: Poor: Some minor overlap exists, but the responses differ substantially on key points or important nuances.
    \item \quad 2: Fair: Several key points overlap, but significant differences remain in multiple important nuances.
    \item \quad 3: Good: Most key points overlap, but the responses differ on one or two minor points.
    \item \quad 4: Excellent: The responses are fully comprehensive, capturing all the same points and nuances.
\end{itemize}

\textbf{DATA TO EVALUATE:} \\
Question: ``\{question\}''\\
\{response 1\}\\
\{response 2\}\\
\\
\textbf{INSTRUCTIONS:}\\
Respond *only* with a JSON object containing:
\begin{enumerate}
    \item ``score": A number from 0 to 4, expressed to one decimal place.
    \item ``justification": A brief explanation for your score (2-4 sentences).
\end{enumerate}

If one response includes information not in the other response, begin your justification with:
\newline
[EXTRA\_INFO]
\newline
If this extra information is also irrelevant to the question, begin your justification with:
\newline
[EXTRA\_INFO] [IRRELEVANT\_INFO]
\newline
 Do not include any text before or after the JSON. 
\end{framed}
\end{figure}

\newpage

\begin{figure}[h]
        \caption{Assessment prompt for LLM judge (GPT-5, high reasoning effort) for LLM to combined expert response comparison (combined overlap score).}
    \label{fig:3v1_assessment_prompt}
    
    \begin{framed}
        You are a medical researcher assessing the ability of LLMs to evaluate and summarise biomedical research papers. An LLM has been given a research paper on the topic of whether HMTV / MMTV-like virus can cause breast cancer in humans, along with a set of questions to answer based on the paper. Score the LLM answer using the scoring rubric below based on its completeness, accuracy, succinctness, and focus compared to the three provided expert reference answers.\\    
        \\
        \textbf{Scoring Rubric:}
        \begin{itemize}
            \item \quad 0: Irrelevant or Incorrect: The generated answer is completely off-topic or factually wrong compared to the references.
            \item \quad 1: Poor: The answer includes a minor point but misses the main idea(s) or important nuances captured in the references.
            \item \quad 2: Fair: The answer captures some of the key ideas but is incomplete, misses some nuances, includes irrelevant information, or is more verbose than necessary.
            \item \quad 3: Good: The answer includes most of the key points from the references but either omits a minor point, includes some irrelevant information, or is slightly more verbose than necessary.
            \item \quad 4: Excellent: The answer is comprehensive, succinct, and does not contain irrelevant information. It is as good as, or better than, the expert reference answers or combination thereof.
\end{itemize}

Here is the data to evaluate:\\        
    Generated Answer:\\       
    \{llm\_response\}\\        
    \{reference\_block\}\\
\\
\textbf{INSTRUCTIONS}: Provide your evaluation in a JSON format with two keys:
    \begin{enumerate}
        \item ``score": An integer from 0 to 4.
        \item ``justification": A brief explanation for your score. If the LLM answer includes any information not present in the references, begin your justification with the flag ``[EXTRA\_INFO]".
    \end{enumerate}
Do not add any text before or after the JSON object.
\end{framed}
\end{figure}

\newpage

\subsection{Tables}

\begin{table}[!htb]
\centering
\begin{threeparttable}
\caption{Manual scoring of pairwise overlap score was done for two question instances (all LLMs included, resulting in 30 scores for comparison). The table shows the number of question instances where the LLM-judge score was an exact match to the human-judge score, and where the LLM-judge score was higher or lower than the human-judge score. The human and LLM-judge scores, along with their respective justifications, can be found in the dataset. Initial human scores were discussed by the research team. The majority of LLM-judge scores were very close to the human-judge scores ($\leq$1 point difference). In total, 10/30 scores were exact matches, 11/30 deviated by $\leq$0.5 points, 7/30 deviated by 1 point, and 2/30 deviated by 2 points.}
\label{tab:1v1_human_score_validation_results}
\begin{tabular}{l >{\centering\arraybackslash}p{2cm} >{\centering\arraybackslash}p{3cm} >{\centering\arraybackslash}p{3cm}}
\toprule
\multicolumn{4}{c}{\textbf{Comparison of human and LLM-judge pairwise overlap scores}} \\
\midrule
\textbf{Evaluated group}&   \textbf{Exact Match}& \textbf{Higher than human judge}&\textbf{Lower than human judge}\\
\midrule
Experts &   1&5\tnote{*}&0\\
\midrule
Gemini 2.5 Pro &   3&1\tnote{**}&2\tnote{**}\\
\midrule
Gemini 2.5 Flash &   3&1\tnote{***}&2\tnote{**}\\
\midrule
GPT-5 &   3&2\tnote{***}&1\tnote{***}\\
\midrule
GPT-5 Nano &   0&5\tnote{***}&1\tnote{***}\\
\midrule
Overall &   10&18&6\\
\bottomrule
\end{tabular}
\begin{tablenotes}
    \item[*] Three deviated by $\leq$0.5 points, while two deviated by 2 points. 
    \item[**] Deviated from human score by $\leq$0.5 points.
    \item[***] Deviated from human score by $\leq$1 point.
\end{tablenotes}
\end{threeparttable}
\end{table}

\begin{table}[!htb]
\centering
\begin{threeparttable}
\caption{Manual scoring of combined overlap score was done for eight question instances (two per LLM). The table shows the number of question instances where the LLM-judge score was an exact match to the human-judge score, and where the LLM-judge score was higher or lower than the human-judge score. The human and LLM-judge scores, along with their respective justifications, can be found in the dataset. Initial human scores were further discussed by the research team. The majority of LLM-judge scores matched the human-judge scores exactly (6/8), with 2/8 scores deviating by 1 point.}
\label{tab:3v1_human_score_validation_results}
\begin{tabular}{l >{\centering\arraybackslash}p{2cm} >{\centering\arraybackslash}p{3cm} >{\centering\arraybackslash}p{3cm}}
\toprule
\multicolumn{4}{c}{\textbf{Comparison of human and LLM-judge combined overlap scores}} \\
\midrule
\textbf{Evaluated LLM}&   \textbf{Exact Match}& \textbf{Higher than human judge}&\textbf{Lower than human judge}\\
\midrule
Gemini 2.5 Pro &   1&1\tnote{*}& 0\\
\midrule
Gemini 2.5 Flash &   1&0& 1\tnote{*}\\
\midrule
GPT-5 &   2&0& 0\\
\midrule
GPT-5 Nano &   2&0& 0\\
\midrule
Overall &   6&1\tnote{*}& 1\tnote{*}\\
\bottomrule
\end{tabular}
\begin{tablenotes}
    \item[*] Deviated from human score by 1 point.
\end{tablenotes}
\end{threeparttable}
\end{table}

\newpage
\subsection{Extraction Template (Questionnaire)}

Category one questions (answered by domain experts) are colored black, while category two questions (Q2-11.1, Q13-16, Q19-21, and Q23-25.1, answered by clinician-scientist trainee) are colored blue. Note that ``SAQ"s in the template are referred to as MCQs in the manuscript, while ``MCQ"s in the template are referred to as multi-select questions in the manuscript.

\noindent\rule{\linewidth}{0.4pt}

You are a medical researcher analysing the literature to evaluate the plausibility of HMTV/MMTV-like virus as a causative agent of breast cancer in humans. For each paper, you will generate a summary and evaluation of its findings by answering 77 predefined questions. These individual summaries will contribute to a final determination of whether HMTV/MMTV-like virus plays a causative role in human breast cancer and whether this warrants further research. To achieve this, carefully review the given paper and respond to questions 1 through 30, some of which are in SAQ or MCQ format, while others are long answer questions.

\textbf{\textit{Relevancy}}

\textbf{1. } How relevant is this article in determining whether it is plausible that HMTV/MMTV-like virus can cause breast cancer in humans (relevant, somewhat relevant, or irrelevant)? A relevant article would fulfil either of the following criteria: 1) gives causal evidence for whether or not HMTV/MMTV-like virus or its species/family of microbes cause breast cancer or other cancers either in humans or animal models, or 2) contains a theoretical model for how HMTV/MMTV-like virus might cause breast cancer, irrespective of whether there is experimental evidence. A somewhat relevant article would fulfil either of the following criteria: 1) mentions both HMTV/MMTV-like virus and breast cancer, but does not give any causal arguments or evidence for HMTV/MMTV-like virus causing breast cancer, or 2) contains evidence examining whether other microbes cause breast cancer, if the findings can plausibly generalise to HMTV/MMTV-like virus and breast cancer. An irrelevant article would be a paper that does not fulfil the criteria for the relevant or somewhat relevant classifications.
\begin{enumerate}[label=\alph*)]
    \item \quad Relevant
    \item \quad Somewhat relevant
    \item \quad Irrelevant
\end{enumerate}

\textbf{1.1 } Briefly explain why the paper is relevant, somewhat relevant, or irrelevant.

Only continue to answer the questions after this if the paper is relevant.

\textbf{\textit{Paper Bibliographic Details}}

\textcolor{blue}{
\textbf{2. } Title of paper}

\textcolor{blue}{
\textbf{3. } Publication date}

\textcolor{blue}{
\textbf{4. } Journal of publication}

\textcolor{blue}{
\textbf{5. } Study design (e.g., randomised control trial, cohort study, case-control study, cross-sectional studies, laboratory study, etc)}

\textcolor{blue}{
\textbf{6. } What country was the research conducted in?}

\textcolor{blue}{
\textbf{7. } DOI}

\textbf{\textit{Paper Integrity \& Reliability}}

\textcolor{blue}{
\textbf{8. } Are the references relevant to the topic of this paper and the sections of the text that they are cited in?
\begin{enumerate}[label=\alph*)]
    \item \quad Yes, all references are relevant.
    \item \quad No, there were irrelevant references.
\end{enumerate}}

\textcolor{blue}{
\textbf{8.1 } If any of the references were irrelevant, list them and state why they are irrelevant.}

\textcolor{blue}{
\textbf{9. } Was there any evidence of conflict of interest in this paper? This may include, but is not limited to, funding of research that could compromise the researchers’ objectivity, and patent ownership related to the research.
\begin{enumerate}[label=\alph*)]
    \item \quad Yes, there was evidence of a conflict of interest in the paper.
    \item \quad No, there was no evidence of a conflict of interest in the paper.
\end{enumerate}}

\textcolor{blue}{
\textbf{9.1 } If there was evidence of a conflict of interest, what was it?}

\textcolor{blue}{
\textbf{10. } Are there any contradictions within the paper? Examples of contradictions include inconsistencies in reported sample sizes and participant characteristics, methodological discrepancies between stated procedures and actual implementation, or variations in how results are reported across different sections of the paper.
\begin{enumerate}[label=\alph*)]
    \item \quad Yes, there were contradictions within the paper.
    \item \quad No, there were no contradictions found within the paper.
\end{enumerate}}

\textcolor{blue}{
\textbf{10.1 } If there were contradictions, list them.}

\textcolor{blue}{
\textbf{11. } Is there any evidence of image or data manipulation in this paper? E.g., using the same image to describe different results by re-labelling, rotating, changing the colours or contrast of the image, or cropping the image differently.
\begin{enumerate}[label=\alph*)]
    \item \quad Yes, there was evidence of image or data manipulation in the paper.
    \item \quad No, there was no evidence of image or data manipulation in the paper.
\end{enumerate}}

\textcolor{blue}{
\textbf{11.1 } If there was evidence of image or data manipulation, what specific alterations were identified?}

\textbf{12. } Determine how reliable this paper is and produce a paper reliability score out of 5.
\begin{enumerate}
    \item[5)] \quad No issues detected.
    \item[4)] \quad Some noticeable issues, such as a few minor contradictions, or an identifiable (but not severe) conflict of interest.
    \item[3)] \quad Integrity is reasonable but weakened by repeated minor contradictions, a few irrelevant citations, or questionable choices in data presentation.
    \item[2)] \quad The paper contains major contradictions, numerous irrelevant references, and/or noticeable conflicts of interest that raise concerns about bias.
    \item[1)] \quad The paper is entirely unreliable due to extensive issues, including irrelevant citations, extreme bias from major conflicts of interest, multiple major contradictions, and/or strong indications of data or image manipulation.
\end{enumerate}

\textbf{\textit{Summary of Paper Contents}}

\textcolor{blue}{
\textbf{13. } What was the aim of the study?}

\textcolor{blue}{
\textbf{14. } What was the hypothesis of the study?}

\textcolor{blue}{
\textbf{15. } Provide a comprehensive, bullet-point summary detailing the demographic characteristics of the study sample, specifically the distribution of age, sex, ethnicity, and economic status.}

\textcolor{blue}{
\textbf{16. } Provide a comprehensive, bullet-point summary detailing the cancer types and subtypes included in the study sample, along with the quantity of each. Specify histological subtypes, grades, stages, receptor status, and any other relevant characteristics.}

\textbf{17. } Based on the sample demographics and the country in which this study was performed, are the findings of this study applicable to African populations in general?
\begin{enumerate}[label=\alph*)]
    \item \quad Yes, the findings are applicable as populations across Africa share similar genetic, socioeconomic, and/or environmental conditions.
    \item \quad Partially applicable, but caution should be exercised due to differences in genetic, socioeconomic, and/or environmental conditions.
    \item \quad No, the findings cannot be applied to African populations due to significant contextual differences.
    \item \quad Applicability cannot be determined without further information about the study’s sample.
\end{enumerate}

\textbf{18. } List the most important factor(s) in the methodology that influence(s) the strength of the study’s findings.

\textcolor{blue}{
\textbf{19. } According to the paper, do the relevant results of this research support any other paper/s published prior to it?
\begin{enumerate}[label=\alph*)]
    \item \quad Yes, this paper’s findings support previous research.
    \item \quad No, this paper’s approach is novel and thus its results do not support or contradict previous findings.
    \item \quad No, this paper’s findings contradict what has previous been found.
    \item \quad Some of the findings support previous research, while other findings contradict previous research.
\end{enumerate}}

\textcolor{blue}{
\textbf{20. } List all the papers referenced in this article that may assist in determining whether it is plausible that HMTV/MMTV-like virus can cause breast cancer in humans. Only include papers that would be classified as relevant as per the original classification criteria.}

\textbf{\textit{Strength of Evidence}}

\textcolor{blue}{
\textbf{21. } If applicable, was the data collected with probability sampling (random selection)?
\begin{enumerate}[label=\alph*)]
    \item \quad Yes, random selection was used.
    \item \quad No, random selection was not used when it should have been.
    \item \quad It is unclear whether random selection was used.
    \item \quad N/A, random selection is not possible with this study design.
\end{enumerate}}

\textbf{22. } If applicable, was the sample size for cases - and, where relevant, controls - sufficient to support the paper’s conclusions?
\begin{enumerate}[label=\alph*)]
    \item \quad Yes, the sample size was sufficient.
    \item \quad The sample size was smaller than ideal, yet sufficiently large that, combined with strong statistical findings, the conclusions remain credible.
    \item \quad No, the sample size was insufficient to support the conclusions.
    \item \quad N/A, no sample used in this study.
\end{enumerate}

\textcolor{blue}{
\textbf{23. } What is the statistical significance of the paper’s relevant findings?}

\textcolor{blue}{
\textbf{24. } Was the statistical analysis appropriate in this paper?
\begin{enumerate}[label=\alph*)]
    \item \quad Yes, all statistical analysis was appropriate.
    \item \quad No, the statistical analysis was inappropriate.
    \item \quad It is unclear what statistical analysis was performed, making it difficult to determine whether it was appropriate.
\end{enumerate}}

\textcolor{blue}{
\textbf{25. } Are there any obvious errors in the statistical analysis or presentation of the paper’s results?
\begin{enumerate}[label=\alph*)]
    \item \quad Yes, there was an error in the paper’s statistical analysis or presentation of results.
    \item \quad No, there were no errors in the paper's statistical analysis or presentation of results.
\end{enumerate}}

\textcolor{blue}{
\textbf{25.1 } If there was an error in the statistical analysis or presentation of the results, what was it?}

\textbf{26. } Are there any limitations to using this research as evidence for or against a plausible causative relationship between HMTV/MMTV-like virus and breast cancer?
\begin{enumerate}[label=\alph*)]
    \item \quad Yes, there are limitations.
    \item \quad No, there are no limitations.
\end{enumerate}

\textbf{26.1 } If there are limitations, what are they?

\textbf{27. } Assess the strength of evidence provided by this paper and assign a score from 1 to 5, with 1 representing severely weak / no credible evidence and 5 representing very strong evidence.
\begin{enumerate}
    \item[5)] \quad Very strong evidence
    \item[4)] \quad Strong evidence
    \item[3)] \quad Acceptable evidence
    \item[2)] \quad Weak evidence
    \item[1)] \quad Severely weak evidence/no credible evidence
\end{enumerate}

\textbf{\textit{Does this paper help fulfil or disprove the microbial oncogenesis criteria?}}

\textbf{28. }Determine if this paper helps fulfil the following listed 10 microbial oncogenicity criteria for the MMTV-like virus/HMTV and breast cancer pair. Give 1 point for each criterion the paper supports or clearly fulfils, 0 for a criterion that is either not examined by the paper or remains uncertain at the end of the paper, and -1 for any criterion that is disproved by this paper. Briefly explain the reasoning for each point or lack thereof given. Finally, determine the cumulative score from these criteria points, allocated as the causal criteria score. When answering these questions, consider the paper reliability score (Q12) and the strength of evidence score (Q30).

 \textbf{28.1 }\textbf{Epidemiologic Association}: Consistent association of a microbe (or a combination of microbes) with a specific cancer type within a population (considering demographics and host factors) of humans (or across all populations), especially when compared to people within the same population(s) without cancer. 
 
 \begin{enumerate}
    \item[a.] \quad Tools to consider:
        \begin{enumerate}[leftmargin=1cm]
            \item[i.] \quad Case-control and Cohort studies
            \item[ii.] \quad Histopathology, immunochemistry, PCR
            \item[iii.] \quad Genomics and mass spectrometry
            \item[iv.] \quad Strong association when consistent findings across different geographical regions and consistent risk ratios in case-control or cohort studies
        \end{enumerate}
\end{enumerate}

\textbf{28.1.1 } Indicate whether the paper supports, refutes, does not investigate, or leaves this criterion uncertain.
\begin{enumerate}[label=\alph*)]
    \item \quad 0, the paper did not examine this criterion.
    \item \quad 0, the paper did investigate this criterion, but the findings are either inconclusive or uncertainty remains due to concerns about the study’s reliability or strength of evidence.
    \item \quad 1, the paper provides evidence that supports this criterion.
    \item \quad -1, the paper provides evidence that refutes this criterion.
\end{enumerate}

\textbf{28.1.2 } If this criterion was supported, refuted, or remains uncertain, select all applicable explanations below that describe why.
\begin{enumerate}[label=\alph*)]
    \item \quad Insufficient power
    \item \quad Inappropriate methodology/tools
    \item \quad Concerns regarding paper reliability
    \item \quad Error/s in the statistical analysis
    \item \quad Sufficient power
    \item \quad Appropriate methodology/tools
    \item \quad Other (specify)
\end{enumerate}

\textbf{28.1.3 } If you selected other, please specify.

\textbf{28.1.4 } State which kind of evidence is provided according to the tools to consider categories and identify the specific finding/s that fulfil or refute this criterion.

 \textbf{28.2 }\textbf{Histopathologic association}: Consistent detection of the microbe (virus, bacterium, fungus, or parasite) in cancer tissues compared to healthy controls
 \begin{enumerate}
    \item[a.] \quad Tools to consider:
        \begin{enumerate}[leftmargin=1cm]
            \item[i.] \quad In Vitro assays (PCR, FACS, imaging)
            \item[ii.] \quad Histopathology, immunochemistry, PCR
            \item[iii.] \quad Genomics and mass spectrometry
        \end{enumerate}
\end{enumerate}

\textbf{28.2.1 } Indicate whether the paper supports, refutes, does not investigate, or leaves this criterion uncertain.
\begin{enumerate}[label=\alph*)]
    \item \quad 0, the paper did not examine this criterion.
    \item \quad 0, the paper did investigate this criterion, but the findings are either inconclusive or uncertainty remains due to concerns about the study’s reliability or strength of evidence.
    \item \quad 1, the paper provides evidence that supports this criterion.
    \item \quad -1, the paper provides evidence that refutes this criterion.
\end{enumerate}

\textbf{28.2.2 } If this criterion was supported, refuted, or remains uncertain, select all applicable explanations below that describe why.
\begin{enumerate}[label=\alph*)]
    \item \quad Insufficient power
    \item \quad Inappropriate methodology/tools
    \item \quad Concerns regarding paper reliability
    \item \quad Error/s in the statistical analysis
    \item \quad Sufficient power
    \item \quad Appropriate methodology/tools
    \item \quad Other (specify)
\end{enumerate}

\textbf{28.2.3 } If you selected other, please specify.

\textbf{28.2.4 } State which kind of evidence is provided according to the tools to consider categories and identify the specific finding/s that fulfil or refute this criterion.

 \textbf{28.3 }\textbf{Temporal association}: Evidence that infection precedes onset of cancer
 \begin{enumerate}
    \item[a.] \quad Tools to consider:
        \begin{enumerate}[leftmargin=1cm]
            \item[i.] \quad Longitudinal (long term) studies
        \end{enumerate}
\end{enumerate}

\textbf{28.3.1 } Indicate whether the paper supports, refutes, does not investigate, or leaves this criterion uncertain.
\begin{enumerate}[label=\alph*)]
    \item \quad 0, the paper did not examine this criterion.
    \item \quad 0, the paper did investigate this criterion, but the findings are either inconclusive or uncertainty remains due to concerns about the study’s reliability or strength of evidence.
    \item \quad 1, the paper provides evidence that supports this criterion.
    \item \quad -1, the paper provides evidence that refutes this criterion.
\end{enumerate}

\textbf{28.3.2 } If this criterion was supported, refuted, or remains uncertain, select all applicable explanations below that describe why.
\begin{enumerate}[label=\alph*)]
    \item \quad Insufficient power
    \item \quad Inappropriate methodology/tools
    \item \quad Concerns regarding paper reliability
    \item \quad Error/s in the statistical analysis
    \item \quad Sufficient power
    \item \quad Appropriate methodology/tools
    \item \quad Other (specify)
\end{enumerate}

\textbf{28.3.3 } If you selected other, please specify.

\textbf{28.3.4 } Identify the specific finding/s that fulfil or refute this criterion.

 \textbf{28.4 }\textbf{Experimental evidence of facilitation of oncogenesis}: Induction of cancer/precancerous changes upon introduction of the isolated microbe into appropriate models. Does oncogenic cellular transformation occur when a microbe is introduced into an animal (ideally a primate or other human-representative) model or tissue model?
\begin{enumerate}
    \item[a.] \quad Tools to consider:
        \begin{enumerate}[leftmargin=1cm]
            \item[i.] \quad 3D biosystems (organoids) or animal models
        \end{enumerate}
\end{enumerate}

\textbf{28.4.1 } Indicate whether the paper supports, refutes, does not investigate, or leaves this criterion uncertain.
\begin{enumerate}[label=\alph*)]
    \item \quad 0, the paper did not examine this criterion.
    \item \quad 0, the paper did investigate this criterion, but the findings are either inconclusive or uncertainty remains due to concerns about the study’s reliability or strength of evidence.
    \item \quad 1, the paper provides evidence that supports this criterion.
    \item \quad -1, the paper provides evidence that refutes this criterion.
\end{enumerate}

\textbf{28.4.2 } If this criterion was supported, refuted, or remains uncertain, select all applicable explanations below that describe why.
\begin{enumerate}[label=\alph*)]
    \item \quad Insufficient power
    \item \quad Inappropriate methodology/tools
    \item \quad Concerns regarding paper reliability
    \item \quad Error/s in the statistical analysis
    \item \quad Sufficient power
    \item \quad Appropriate methodology/tools
    \item \quad Other (specify)
\end{enumerate}

\textbf{28.4.3 } If you selected other, please specify.

\textbf{28.4.4 } Identify the specific finding/s that fulfil or refute this criterion.

\textbf{28.5 }\textbf{Molecular and Multi-omics evidence for interaction}:
\begin{enumerate}
    \item[i.] \quad Epigenetics (e.g. DNA methylation changes in H. pylori or HPV E6 oncogene resulting in P53 degradation), as well as mutational signatures
    \item[ii.] \quad Mutation effects (stimulating cellular mutations promoting cancer)
    \item[iii.] \quad Stimulating immune evasion mechanisms reducing immunosurveillance for emerging cancer cells
    \item[iv.] \quad Integration of microbial DNA into host genome 
\end{enumerate}
\begin{enumerate}
    \item[a.] \quad Tools to consider:
        \begin{enumerate}[leftmargin=1cm]
            \item[i.] \quad in vitro assays, organoids, multi-omics assessments including:
                \begin{enumerate}[leftmargin=1cm]
                    \item[a.] \quad \textbf{Molecular Evidence}
                    \begin{enumerate}[leftmargin=1cm]
                        \item[i.] \quad For bacteria: Identification of toxins, effector proteins, or metabolites that alter cellular signalling
                        \item[ii.] \quad For viruses: Characterization of viral oncoproteins or insertional mutagenesis
                        \item[iii.] \quad For fungi: Demonstration of mycotoxins or immune-modulating molecules with carcinogenic potential
                        \item[iv.] \quad For parasites: Identification of chronic inflammatory responses or direct tissue damage mechanisms
                        \item[v.] \quad Evidence of microbial interference with DNA repair, cell cycle regulation, apoptosis, or immune surveillance
                    \end{enumerate}
                    \item[b.] \quad \textbf{Genomic evidence:}
                    \begin{enumerate}[leftmargin=1cm]
                        \item[i.] \quad Microbial DNA sequences or integration sites in tumor genome; 
                        \item[ii.] \quad Host genetic susceptibility factors that enhance oncogenic potential of specific microbes, including demonstration of how microbial exposure can alter (or promote) known host genetic risk factors for cancer.
                        \item[iii.] \quad Demonstratable facilitation of cancer-associated mutational signatures.
                    \end{enumerate}
                    \item[c.] \quad \textbf{Transcriptomic evidence:}
                    \begin{enumerate}[leftmargin=1cm]
                        \item[i.] \quad Expression of microbial genes or altered host gene expression profiles
                    \end{enumerate}
                    \item[d.] \quad \textbf{Proteomic evidence:}
                    \begin{enumerate}[leftmargin=1cm]
                        \item[i.] \quad  Detection of microbial proteins or altered host protein responses
                    \end{enumerate}
                    \item[e.] \quad \textbf{Metabolomic evidence:}
                    \begin{enumerate}[leftmargin=1cm]
                        \item[i.] \quad Microbial metabolites or altered host metabolic pathways
                    \end{enumerate}
                    \item[f.] \quad \textbf{Microbiome analyses:}
                    \begin{enumerate}[leftmargin=1cm]
                        \item[i.] \quad Consistent dysbiosis patterns associated with specific cancers
                    \end{enumerate}
                \end{enumerate}
        \end{enumerate}
\end{enumerate}

\textbf{28.5.1 } Indicate whether the paper supports, refutes, does not investigate, or leaves this criterion uncertain.
\begin{enumerate}[label=\alph*)]
    \item \quad 0, the paper did not examine this criterion.
    \item \quad 0, the paper did investigate this criterion, but the findings are either inconclusive or uncertainty remains due to concerns about the study’s reliability or strength of evidence.
    \item \quad 1, the paper provides evidence that supports this criterion.
    \item \quad -1, the paper provides evidence that refutes this criterion.
\end{enumerate}

\textbf{28.5.2 } If this criterion was supported, refuted, or remains uncertain, select all applicable explanations below that describe why.
\begin{enumerate}[label=\alph*)]
    \item \quad Insufficient power
    \item \quad Inappropriate methodology/tools
    \item \quad Concerns regarding paper reliability
    \item \quad Error/s in the statistical analysis
    \item \quad Sufficient power
    \item \quad Appropriate methodology/tools
    \item \quad Other (specify)
\end{enumerate}

\textbf{28.5.3 } If you selected other, please specify.

\textbf{28.5.4 } State which kind of evidence is provided according to the tools to consider categories and identify the specific finding/s that fulfil or refute this criterion.

 \textbf{28.6 }\textbf{Prevention}: Does preventing the infection (or removing exposure to the microbe) reduce cancer or evidence of oncogenesis
\begin{enumerate}
    \item[a.] \quad Tools to consider:
        \begin{enumerate}[leftmargin=1cm]
            \item[i.] \quad Clinical trials
            \item[ii.] \quad Relevant animal models
            \item[iii.] \quad For viruses: Reduced cancer incidence following vaccination (e.g., HPV, HBV)
            \item[iv.] \quad For bacteria: Cancer prevention through antibiotic treatment or bacterial elimination
            \item[v.] \quad For fungi: Antifungal intervention effects on precancerous lesions
            \item[vi.] \quad For parasites: Impact of antiparasitic treatment on cancer development
            \item[vii.] \quad Prevention trials showing reduced cancer incidence after targeting the microbe
        \end{enumerate}
\end{enumerate}

\textbf{28.6.1 } Indicate whether the paper supports, refutes, does not investigate, or leaves this criterion uncertain.
\begin{enumerate}[label=\alph*)]
    \item \quad 0, the paper did not examine this criterion.
    \item \quad 0, the paper did investigate this criterion, but the findings are either inconclusive or uncertainty remains due to concerns about the study’s reliability or strength of evidence.
    \item \quad 1, the paper provides evidence that supports this criterion.
    \item \quad -1, the paper provides evidence that refutes this criterion.
\end{enumerate}

\textbf{28.6.2 } If this criterion was supported, refuted, or remains uncertain, select all applicable explanations below that describe why.
\begin{enumerate}[label=\alph*)]
    \item \quad Insufficient power
    \item \quad Inappropriate methodology/tools
    \item \quad Concerns regarding paper reliability
    \item \quad Error/s in the statistical analysis
    \item \quad Sufficient power
    \item \quad Appropriate methodology/tools
    \item \quad Other (specify)
\end{enumerate}

\textbf{28.6.3 } If you selected other, please specify.

\textbf{28.6.4 } State which kind of evidence is provided according to the tools to consider categories and identify the specific finding/s that fulfil or refute this criterion.

\textbf{28.7 }\textbf{Reproducibility and Validation}: Confirmation across independent laboratories, optimally using varied methodologies or models. Replication in diverse human populations and geographic settings. Concordance between in vitro, animal, and human studies

\textbf{28.7.1 } Indicate whether the paper supports, refutes, does not investigate, or leaves this criterion uncertain.
\begin{enumerate}[label=\alph*)]
    \item \quad 0, the paper did not examine this criterion.
    \item \quad 0, the paper did investigate this criterion, but the findings are either inconclusive or uncertainty remains due to concerns about the study’s reliability or strength of evidence.
    \item \quad 1, the paper provides evidence that supports this criterion.
    \item \quad -1, the paper provides evidence that refutes this criterion.
\end{enumerate}

\textbf{28.7.2 } If this criterion was supported, refuted, or remains uncertain, select all applicable explanations below that describe why.
\begin{enumerate}[label=\alph*)]
    \item \quad Insufficient power
    \item \quad Inappropriate methodology/tools
    \item \quad Concerns regarding paper reliability
    \item \quad Error/s in the statistical analysis
    \item \quad Sufficient power
    \item \quad Appropriate methodology/tools
    \item \quad Other (specify)
\end{enumerate}

\textbf{28.7.3 } If you selected other, please specify.

\textbf{28.7.4 } Identify the specific finding/s that fulfil or refute this criterion.

\textbf{28.8 }\textbf{Dose Response relationship}: Relationship of severity or chronicity of the presumed offending infection with cancer initiation and severity in human longitudinal studies or in experimental models demonstrate the association of infectious dose with development of cancer
\begin{enumerate}
    \item[a.] \quad Tools to consider:
        \begin{enumerate}[leftmargin=1cm]
            \item[i.] \quad Organoids, animal models, or clinico-epidemiologic longitudinal studies
        \end{enumerate}
\end{enumerate}

\textbf{28.8.1 } Indicate whether the paper supports, refutes, does not investigate, or leaves this criterion uncertain.
\begin{enumerate}[label=\alph*)]
    \item \quad 0, the paper did not examine this criterion.
    \item \quad 0, the paper did investigate this criterion, but the findings are either inconclusive or uncertainty remains due to concerns about the study’s reliability or strength of evidence.
    \item \quad 1, the paper provides evidence that supports this criterion.
    \item \quad -1, the paper provides evidence that refutes this criterion.
\end{enumerate}

\textbf{28.8.2 } If this criterion was supported, refuted, or remains uncertain, select all applicable explanations below that describe why.
\begin{enumerate}[label=\alph*)]
    \item \quad Insufficient power
    \item \quad Inappropriate methodology/tools
    \item \quad Concerns regarding paper reliability
    \item \quad Error/s in the statistical analysis
    \item \quad Sufficient power
    \item \quad Appropriate methodology/tools
    \item \quad Other (specify)
\end{enumerate}

\textbf{28.8.3 } If you selected other, please specify.

\textbf{28.8.4 } State which kind of evidence is provided according to the tools to consider categories and identify the specific finding/s that fulfil or refute this criterion.

\textbf{28.9 }\textbf{Plausibility}: Are there plausible mechanisms for considering a potential role for a microbe
\begin{enumerate}
    \item[a.] \quad Tools to consider:
        \begin{enumerate}[leftmargin=1cm]
            \item[i.] \quad Knowledge-based; i.e. understanding the physiology of microbial interaction with humans may suggest or support a role in carcinogenesis
        \end{enumerate}
\end{enumerate}

\textbf{28.9.1 } Indicate whether the paper supports, refutes, does not discuss, or leaves this criterion uncertain.
\begin{enumerate}[label=\alph*)]
    \item \quad 0, the paper does not suggest or discuss any plausible mechanisms for the MCP.
    \item \quad 0, the paper suggests a possible mechanism, but it is weakly justified, inconsistent with existing knowledge, or supported by insufficient evidence from the study.
    \item \quad 1, the paper presents a plausible mechanism, either consistent with existing knowledge of cancer biology or microbial pathogenesis or supported by credible evidence produced by the present study.
    \item \quad -1, the paper presents reasoning or mechanisms that argue against a plausible role for the microbe in carcinogenesis.
\end{enumerate}

\textbf{28.9.2 } If this criterion was supported, refuted, or remains uncertain, select all applicable explanations below that describe why.
\begin{enumerate}[label=\alph*)]
    \item \quad Insufficient power
    \item \quad Inappropriate methodology/tools
    \item \quad Concerns regarding paper reliability
    \item \quad Error/s in the statistical analysis
    \item \quad Sufficient power
    \item \quad Appropriate methodology/tools
    \item \quad Other (specify)
\end{enumerate}

\textbf{28.9.3 } If you selected other, please specify.

\textbf{28.9.4 } Identify the specific finding/s that fulfil or refute this criterion.

 \textbf{28.10 }\textbf{Impact of co-Factors}: Microbial carcinogenesis occurs (or is accelerated) in an environment with promotive host factors, such as genetic risk, immunodeficiencies, nutritional status, and/or with environmental factors (including but not limited to known carcinogens)
\begin{enumerate}
    \item[a.] \quad Tools to consider:
        \begin{enumerate}[leftmargin=1cm]
            \item[i.] \quad Epidemiological studies in humans
            \item[ii.] \quad Laboratory studies
            \item[iii.] \quad Animal models
        \end{enumerate}
\end{enumerate}

\textbf{28.10.1 } Indicate whether the paper supports, refutes, does not investigate, or leaves this criterion uncertain.
\begin{enumerate}[label=\alph*)]
    \item \quad 0, the paper did not examine this criterion.
    \item \quad 0, the paper did investigate this criterion, but the findings are either inconclusive or uncertainty remains due to concerns about the study’s reliability or strength of evidence.
    \item \quad 1, the paper provides evidence that supports this criterion.
    \item \quad -1, the paper provides evidence that refutes this criterion.
\end{enumerate}

\textbf{28.10.2 } If this criterion was supported, refuted, or remains uncertain, select all applicable explanations below that describe why.
\begin{enumerate}[label=\alph*)]
    \item \quad Insufficient power
    \item \quad Inappropriate methodology/tools
    \item \quad Concerns regarding paper reliability
    \item \quad Error/s in the statistical analysis
    \item \quad Sufficient power
    \item \quad Appropriate methodology/tools
    \item \quad Other (specify)
\end{enumerate}

\textbf{28.10.3 } If you selected other, please specify.

\textbf{28.10.4 } State which kind of evidence is provided according to the tools to consider categories and identify the specific finding/s that fulfil or refute this criterion.

 \textbf{29 } Calculate the cumulative score from these criteria points and provide the final causal criteria score.

 \textbf{30 } Indicate whether the paper supports, refutes, or does not impact the plausibility of HMTV/MMTV-like virus causing breast cancer.
\begin{enumerate}
    \item[5)] \quad Strongly supports (moderate to strong evidence)
    \item[4)] \quad Weakly supports (weak to moderate evidence)
    \item[3)] \quad Does not impact (negligible or weak evidence)
    \item[2)] \quad Weakly refutes (weak to moderate evidence)
    \item[1)] \quad Strongly refutes (moderate to strong evidence)
\end{enumerate}

\newpage
\subsection{Abridged Extraction Template for Repeat Experiments}
This version of the extraction template was used for repeat experiments (30) with GPT-5. For the second paper, question item 4.4 (microbial oncogenesis criterion ``Impact of co-factors") was replaced with the microbial oncogenesis criterion ``Plausibility" question item (see question item 28.9 and 28.9.1 in the full extraction template).

\noindent\rule{\linewidth}{0.4pt}

You are a medical researcher analysing the literature to evaluate the plausibility of HMTV/MMTV-like virus as a causative agent of breast cancer in humans. To achieve this, carefully review the given paper and respond to the questions that follow.

\textbf{\textit{Relevancy}}

\textbf{1. } How relevant is this article in determining whether it is plausible that HMTV/MMTV-like virus can cause breast cancer in humans (relevant, somewhat relevant, or irrelevant)? A relevant article would fulfil either of the following criteria: 1) gives causal evidence for whether or not HMTV/MMTV-like virus or its species/family of microbes cause breast cancer or other cancers either in humans or animal models, or 2) contains a theoretical model for how HMTV/MMTV-like virus might cause breast cancer, irrespective of whether there is experimental evidence. A somewhat relevant article would fulfil either of the following criteria: 1) mentions both HMTV/MMTV-like virus and breast cancer, but does not give any causal arguments or evidence for HMTV/MMTV-like virus causing breast cancer, or 2) contains evidence examining whether other microbes cause breast cancer, if the findings can plausibly generalise to HMTV/MMTV-like virus and breast cancer. An irrelevant article would be a paper that does not fulfil the criteria for the relevant or somewhat relevant classifications.
\begin{enumerate}[label=\alph*)]
    \item \quad Relevant
    \item \quad Somewhat relevant
    \item \quad Irrelevant
\end{enumerate}

\textbf{\textit{Strength of Evidence}}

\textbf{2. } If applicable, was the sample size for cases - and, where relevant, controls - sufficient to support the paper’s conclusions?
\begin{enumerate}[label=\alph*)]
    \item \quad Yes, the sample size was sufficient.
    \item \quad The sample size was smaller than ideal, yet sufficiently large that, combined with strong statistical findings, the conclusions remain credible.
    \item \quad No, the sample size was insufficient to support the conclusions.
    \item \quad N/A, no sample used in this study.
\end{enumerate}

\textbf{3. } Assess the strength of evidence provided by this paper and assign a score from 1 to 5, with 1 representing severely weak / no credible evidence and 5 representing very strong evidence.
\begin{enumerate}
    \item[5)] \quad Very strong evidence
    \item[4)] \quad Strong evidence
    \item[3)] \quad Acceptable evidence
    \item[2)] \quad Weak evidence
    \item[1)] \quad Severely weak evidence/no credible evidence
\end{enumerate}

\textbf{\textit{Does this paper help fulfil or disprove the microbial oncogenesis criteria?}}

\textbf{4. }Determine if this paper helps fulfil the following listed microbial oncogenicity criteria for the MMTV-like virus/HMTV and breast cancer pair. Give 1 point for each criterion the paper supports or clearly fulfils, 0 for a criterion that is either not examined by the paper or remains uncertain at the end of the paper, and -1 for any criterion that is disproved by this paper. When answering these questions, consider the strength of evidence score (Q3).

 \textbf{4.1 }\textbf{Epidemiologic Association}: Consistent association of a microbe (or a combination of microbes) with a specific cancer type within a population (considering demographics and host factors) of humans (or across all populations), especially when compared to people within the same population(s) without cancer. 
 
 \begin{enumerate}
    \item[a.] \quad Tools to consider:
        \begin{enumerate}[leftmargin=1cm]
            \item[i.] \quad Case-control and Cohort studies
            \item[ii.] \quad Histopathology, immunochemistry, PCR
            \item[iii.] \quad Genomics and mass spectrometry
            \item[iv.] \quad Strong association when consistent findings across different geographical regions and consistent risk ratios in case-control or cohort studies
        \end{enumerate}
\end{enumerate}

\textbf{4.1.1 } Indicate whether the paper supports, refutes, does not investigate, or leaves this criterion uncertain.
\begin{enumerate}[label=\alph*)]
    \item \quad 0, the paper did not examine this criterion.
    \item \quad 0, the paper did investigate this criterion, but the findings are either inconclusive or uncertainty remains due to concerns about the study’s reliability or strength of evidence.
    \item \quad 1, the paper provides evidence that supports this criterion.
    \item \quad -1, the paper provides evidence that refutes this criterion.
\end{enumerate}

 \textbf{4.2 }\textbf{Histopathologic association}: Consistent detection of the microbe (virus, bacterium, fungus, or parasite) in cancer tissues compared to healthy controls
 \begin{enumerate}
    \item[a.] \quad Tools to consider:
        \begin{enumerate}[leftmargin=1cm]
            \item[i.] \quad In Vitro assays (PCR, FACS, imaging)
            \item[ii.] \quad Histopathology, immunochemistry, PCR
            \item[iii.] \quad Genomics and mass spectrometry
        \end{enumerate}
\end{enumerate}

\textbf{4.2.1 } Indicate whether the paper supports, refutes, does not investigate, or leaves this criterion uncertain.
\begin{enumerate}[label=\alph*)]
    \item \quad 0, the paper did not examine this criterion.
    \item \quad 0, the paper did investigate this criterion, but the findings are either inconclusive or uncertainty remains due to concerns about the study’s reliability or strength of evidence.
    \item \quad 1, the paper provides evidence that supports this criterion.
    \item \quad -1, the paper provides evidence that refutes this criterion.
\end{enumerate}

\textbf{4.3 }\textbf{Dose Response relationship}: Relationship of severity or chronicity of the presumed offending infection with cancer initiation and severity in human longitudinal studies or in experimental models demonstrate the association of infectious dose with development of cancer
\begin{enumerate}
    \item[a.] \quad Tools to consider:
        \begin{enumerate}[leftmargin=1cm]
            \item[i.] \quad Organoids, animal models, or clinico-epidemiologic longitudinal studies
        \end{enumerate}
\end{enumerate}

\textbf{4.3.1 } Indicate whether the paper supports, refutes, does not investigate, or leaves this criterion uncertain.
\begin{enumerate}[label=\alph*)]
    \item \quad 0, the paper did not examine this criterion.
    \item \quad 0, the paper did investigate this criterion, but the findings are either inconclusive or uncertainty remains due to concerns about the study’s reliability or strength of evidence.
    \item \quad 1, the paper provides evidence that supports this criterion.
    \item \quad -1, the paper provides evidence that refutes this criterion.
\end{enumerate}

 \textbf{4.4 }\textbf{Impact of co-Factors}: Microbial carcinogenesis occurs (or is accelerated) in an environment with promotive host factors, such as genetic risk, immunodeficiencies, nutritional status, and/or with environmental factors (including but not limited to known carcinogens)
\begin{enumerate}
    \item[a.] \quad Tools to consider:
        \begin{enumerate}[leftmargin=1cm]
            \item[i.] \quad Epidemiological studies in humans
            \item[ii.] \quad Laboratory studies
            \item[iii.] \quad Animal models
        \end{enumerate}
\end{enumerate}

\textbf{4.4.1 } Indicate whether the paper supports, refutes, does not investigate, or leaves this criterion uncertain.
\begin{enumerate}[label=\alph*)]
    \item \quad 0, the paper did not examine this criterion.
    \item \quad 0, the paper did investigate this criterion, but the findings are either inconclusive or uncertainty remains due to concerns about the study’s reliability or strength of evidence.
    \item \quad 1, the paper provides evidence that supports this criterion.
    \item \quad -1, the paper provides evidence that refutes this criterion.
\end{enumerate}

 \textbf{5 } Indicate whether the paper supports, refutes, or does not impact the plausibility of HMTV/MMTV-like virus causing breast cancer.
\begin{enumerate}
    \item[5)] \quad Strongly supports (moderate to strong evidence)
    \item[4)] \quad Weakly supports (weak to moderate evidence)
    \item[3)] \quad Does not impact (negligible or weak evidence)
    \item[2)] \quad Weakly refutes (weak to moderate evidence)
    \item[1)] \quad Strongly refutes (moderate to strong evidence)
\end{enumerate}

\end{document}